\documentstyle[prb,aps]{revtex}

\begin{document}


\title{Spin Instabilities in Coupled Semiconductor Quantum Wells}
\author{P. I. Tamborenea,\cite{paddress} R. J. Radtke, and S. Das Sarma}
\address{Condensed Matter Theory Group, Department of Physics,\\
University of Maryland, College Park, Maryland  20742-4111}
\date{\today}
\maketitle

\begin{abstract}
We study the magnetic phases of two coupled two-dimensional electron
gases in order to determine under what circumstances these
phases may occur in real semiconductor quantum wells and what the
experimental properties of the broken-symmetry ground states may be.
Within the local-density-approximation to time-dependent density functional
theory (DFT), we find a phase transition signaled by the vanishing
of the intersubband spin-density excitations at low but accessible
($\sim 10^{10}-10^{11}$ cm$^{-2}$) electron densities.
Through a self-consistent Hartree-Fock calculation, we associate this
transition with an antiferromagnetic phase and study the
phase diagram, thermodynamics, and collective modes in it.
The collective modes are in principle observable in inelastic light
scattering experiments, and we discuss the implications of our calculations
for these measurements.
We also examine the ferromagnetic transition in both single and double
quantum wells within the local-spin-density approximation to DFT
and obtain a critical density which depends on the well width and
which is far below that of the antiferromagnetic transition.
\end{abstract}




\section{INTRODUCTION}
\label{sec:introduction}

The study of spin instabilities holds a fascination for both
theoretical and experimental condensed matter physicists.
To theorists, these instabilities illustrate the qualitatively new
states of matter which may result in simple systems through the
presence of electron-electron interactions.
To experimentalists, they lead to novel phases with unique and
potentially useful properties.
Consequently, the search for systems which exhibit new or unusual spin
instabilities is an area of active research, and any guidance that
theory can provide which suggests avenues for this investigation
should be welcome.

One class of systems ready for serious exploration are single- or
double-quantum-well structures at low density.
Theoretically, the low-dimensionality of these structures restricts
the phase space available for electron-electron scattering,
increasing the relative importance of the interaction and thereby
enhancing the potential for novel phase transitions.
Another, more practical, reason for looking at these structures is that
they can be physically realized in semiconductor space-charge layers and,
in particular, ultra-pure modulation-doped GaAs/Al$_x$Ga$_{1-x}$As
heterostructures.
These devices may be fabricated with precisely
controlled dimensions that are nearly free of defects and which
are tunable over a wide range of densities and band structures.
This freedom yields a large parameter space in which interesting
effects may be found and explored with both experimental and
theoretical tools.

In isolating the interesting regions of this parameter space,
researchers can be guided by simple energetic considerations.
When the kinetic energy of an electron gas dominates the Coulomb
repulsion, as at high electron densities, the system behaves like
a nearly ideal Fermi gas.
Therefore, novel phases will be found only in the
regime in which the kinetic energy is smaller than or of the
same order as the potential energy.
Two ways of reaching this limit suggest themselves:
applying a magnetic field or reducing the electron density.
In a strong magnetic field, the kinetic energy is quenched by the
Landau quantization, leading to a variety of strongly correlated
quantum phases, the best-known example being the fractional quantum Hall
liquid.\cite{tsu-sto-gos,lau,fer,boe,eis,pri-pla-he,he-das-xie,hal-lee-rea}
At extremely low densities, the electrons prefer to crystallize,
as predicted some time ago by Wigner.\cite{wig}
Experimental evidence for this crystallization has been somewhat
equivocal in semiconductor heterostructures, but it is clearly seen in
a two-dimensional electron gas suspended above the surface of
liquid helium.\cite{Grimes}
For somewhat higher densities, magnetic instabilities to spin-density
wave\cite{spi-den-wav} or ferromagnetic\cite{blo} phases have been
proposed.

The parameter space defined by the energetics of a single
low-dimensional electron gas is well defined and thoroughly explored.
To find new effects, many current investigations of these systems
add an additional degree of freedom by coupling two two-dimensional
electron gases together.
The resulting two-layer system has an additional energy scale due to the
splitting of the isolated quantum well levels into symmetric and
antisymmetric components which competes with the intra- and inter-layer
Coulomb energies.
In a high magnetic field, this additional degree of freedom
can lead to the disappearance of odd-integer and the appearance of
even-integer fractional quantized Hall steps\cite{boe,eis,he-das-xie}
and possibly to spontaneous charge transfer between the
layers.\cite{MacDonald,Varma}

In zero-field, a number of experimental results
and theoretical predictions have also appeared in the
literature.\cite{pinczuk,shanabrook,Ruden,nei,mar-das,Katayama,decca,tam-das-94-3,ern-gon-sya,das-tam-94-4,rad-das}
Many of these zero-field studies focus on the behavior of the
collective excitations involving transitions between the
symmetric and antisymmetric levels and which are therefore
unique to two-layer systems.
These excitations can be produced in either the charge or the spin
channels and have energies which are sensitive to the many-body
interactions in the system.
For example, in quantum wells with either square or parabolic confining
potentials, theoretical work suggests that there exists a critical density
below which the many-body corrections to the charge-density excitations
cause the energy of this mode to drop below the energy of the
symmetric-antisymmetric splitting.\cite{mar-das}
This effect was recently observed by inelastic light 
scattering experiments.\cite{ern-gon-sya}
Another experimental study of coupled double-quantum-well
systems reveals that the spin-density excitation energy abruptly merges
with the continuum of intersubband single-particle excitations
as the electron density is increased beyond the point at which the
second-lowest subband begins to populate.\cite{decca}
Time-dependent local-density approximation calculations\cite{tam-das-94-3}
agree quite well with this experiment for most of the range of
densities used, but the abrupt merge seems to demand a more
refined calculation.

Of most relevance to the subject of this paper, recent calculations
of the intersubband spin-density excitations in coupled
double-quantum-well systems indicate that the energy of the lowest
intersubband spin-density excitation may vanish at sufficiently
low density.\cite{das-tam-94-4}
The obvious interpretation of this collapse is that it indicates
an electronic phase transition from the metallic Fermi-liquid phase
to a condensate of zero-energy spin-density excitations.
Since these excitations involve an intersubband electronic transition
accompanied by a spin-flip, this condensate has been termed a
spin-triplet intersubband exciton liquid.
The word ``exciton'' here does not refer to the usual
bound state of an electron in the conduction band and a hole in the
valence band of the semiconductor;
rather, it is used as a reminder that the final-state interaction
or vertex correction is included in the calculation of the
spin-density excitations.\cite{exciton-note}
In general terms, this transition involves the electronic spin
in a fundamental way, suggesting that the new ground state would have
non-trivial magnetic properties.
Hence, the region of parameter space including the quantum well
structures exhibiting this spin-density excitation collapse are
ideal candidates for the study of novel spin instabilities.

In this paper, we examine the question of spin instabilities
in such single- and double-quantum-well structures in the absence of
an external magnetic field.
Our goal is to determine in what structures and under what conditions
these instabilities may occur and what the experimental signatures
of the new ground states may be.
Our primary interest is in the spin-density-excitation-collapsed
phase discussed above,\cite{das-tam-94-4} but we also explore the more
general question of ferromagnetism in single- and double-quantum-well
structures.
We predict that the SDE-softened phase will occur in fairly typical
coupled double-quantum-well structures at low but accessible densities of
order $10^{10}-10^{11}$ cm$^{-2}$, and we show that this phase corresponds
to antiferromagnetic order of the spin densities in the two wells.
By constructing a minimal model of the antiferromagnetic state and
treating it within mean-field theory, we are able to discuss the
qualitative features of this phase.
In particular, we find that the transition to this state may occur at
temperatures on the order of the splitting between the lowest two subbands
in the quantum wells, which can be around 10~K.
Moreover, although transport measurements will likely show no pronounced
anomaly at the transition, both the collective excitations and
the specific heat show distinctive features which can be used to
identify the antiferromagnetic phase.
Lowering the electron density further, our calculations indicate that
these systems re-enter the normal state and then enter a ferromagnetic
phase at densities around $10^{9}$ cm$^{-2}$.
In wide single-quantum-well structures as well, both ferro- and
antiferromagnetic phases exist, and we study the critical
density of the ferromagnetic transition as a function of the well width.
As expected from the increasing importance of exchange effects in
lower dimensions, we find that the ferromagnetic phase is stable
below a critical density which increases as the well width decreases.

This paper is structured in the following way.
In Sec.~\ref{sec:formalism}, we discuss the formalism which underlies
our computations.
In the normal and ferromagnetic states, this formalism is density
functional theory in the local- (LDA) and local-spin-density approximations
(LSDA), respectively.
A time-dependent version of the LDA is also reviewed in Sec.~\ref{sec:LDA},
which provides quantitative results for the collective excitation
spectra that we discuss in Sec.~\ref{sec:instability}.
Additionally, we describe in Sec.~\ref{sec:SCHF} the equations for the
self-energy and the density response function in a self-consistent
Hartree-Fock theory, which is the basis of our calculations within the
antiferromagnetic state.
In Sec.~\ref{sec:instability}, we reproduce and extend the results
of Ref.~\onlinecite{das-tam-94-4}, which predicted the softening of the
intersubband spin-density excitations, and characterize the phase
transition on the paramagnetic side in greater detail.
Section~\ref{sec:FM} contains a study of the ferromagnetic transition in
these systems within the LSDA, and we demonstrate that
that the instability predicted in Sec.~\ref{sec:instability}
occurs at much higher density than the ferromagnetic transition
and so cannot be associated with it.
We also take a first step in the study of the ground-state
spin-polarization of the inhomogeneous electron gas in semiconductor
quantum wells by calculating the critical density of the ferromagnetic
transition in square single quantum wells as a function of well width
at zero temperature.
In Sec.~\ref{sec:AF}, we develop a simple model for the spin-density
instability of Sec.~\ref{sec:instability} and treat it within the
self-consistent Hartree-Fock theory of Sec.~\ref{sec:SCHF}.
We study the ground state, thermodynamic quantities, and the collective
mode spectrum in the broken-symmetry phase and discuss their experimental
ramifications.
Section~\ref{sec:conclusion} presents some speculations on the importance
of non-trivial spin-density modulations transverse to the layering direction
in these quantum well structures and summarizes the results of this paper.

\section{FORMALISM}
\label{sec:formalism}

\subsection{Density-Functional Theory}
\label{sec:LDA}

\subsubsection{Unpolarized Electron Gas (LDA)}
\label{sec:LDA-unpolarized}

The central aim of this Subsection is to develop the formalism used to
compute the intersubband collective excitations in the unpolarized
state of double and wide single quantum wells that we will use in
Sec.~\ref{sec:instability}.
We compute these collective excitations within the so-called
time-dependent local-density approximation (TDLDA).
This approach was first employed by Ando\cite{ando76} to compute
intersubband charge-density excitations (CDEs) and later extended by
Katayama and Ando\cite{kat-and} to study resonant inelastic light
scattering in semiconductor structures.
The use of inelastic light scattering in these systems is motivated
by the fact the charge- and the spin-density excitations (SDEs) couple
to the light polarization differently, and this fact allows a selective
measurement of both types of collective modes.\cite{pin-abs,das:ils}
Detailed descriptions of the TDLDA method for calculating CDE and
SDE energies and spectra have been given in the
literature.\cite{kat-and,mar-das}
However, for the sake of completeness and to facilitate the discussion
of our results in Sec.~\ref{sec:instability}, we describe the TDLDA
approach in some detail in the remainder of this Subsection.
In the next Subsection, we shall generalize this formalism to allow
for the possibility of a spin-polarized ground state of the electron gas.

The first step in the TDLDA calculation of the intersubband excitations
consists of obtaining the renormalized subband energies in the 
local-density approximation (LDA) of Hohenberg, Kohn, and
Sham.\cite{hoh-koh,koh-sha,ste-das}
We choose a coordinate system with the $z$-axis is along the
direction of confinement in the quantum well structure.
The effective single-particle Schr\"{o}dinger equation or Kohn-Sham
equation for this system then reads
\begin{equation}
\left(-\frac{\hbar^2}{2m^*} \nabla^2 + V_{EFF} (z) \right)
  \Psi({\bf R})
  = E \Psi ({\bf R}),
\end{equation}
where we have assumed that the effective electron mass $m^*$ is
constant across the well, ${\bf R} = ({\bf r},z)$ denotes a
three-dimensional vector, and the self-consistent effective potential
$V_{EFF}(z)$ is given below.
The in-plane ($xy$) and $z$-dependences can be separated, and, due to the
assumed translational invariance in the $xy$ plane (the localized donor 
charges are assumed to be smeared out uniformly in the plane), the 
eigenenergies and wave functions become
\begin{equation}
E_{n{\bf k}}=\varepsilon_n + \frac{\hbar^2 k^2}{2m^*}
\end{equation}
and
\begin{equation}
\Psi_{n {\bf k}} ({\bf R}) =
  \frac{1}{\sqrt{A}} e^{i{\bf k \cdot r}} \,\phi_n (z).
\label{eq:factorize}
\end{equation}
In these expressions, $A$ is the sample area,
${\bf k}$ is the in-plane wave vector of the
electron, and $\varepsilon_n$ and $\phi_n (z)$ are the solutions to the
one-dimensional Kohn-Sham equation
\begin{equation}
\left(-\frac{\hbar^2}{2m^*} \frac{d^2}{dz^2} + V_{EFF}(z) \right) \phi_n(z)
  = \varepsilon_n \phi_n(z).
\label{eq:kohn-sham}
\end{equation}

The effective single-particle potential
\begin{equation}
V_{EFF}( z) = V_{CONF} (z) + V_H (z) + V_{XC}( z)
\label{eq:eff-pot}
\end{equation}
contains the confining potential of the bare quantum well
$V_{CONF}(z)$, and the self-consistent Hartree and exchange-correlation
potentials $V_H(z)$ and $V_{XC}(z)$, respectively.
The Hartree potential takes into account the average electrostatic
interaction with the other electrons and the positively charged
donor ions, and is given by the solution of the Poisson equation
\begin{equation}
\frac{d^2V_H(z)}{dz^2}=-\frac{4 \pi e^2}{\epsilon}[n(z)-N_D(z)],
  \label{eq:poisson}
\end{equation}
where $\epsilon$ is the static dielectric constant of GaAs,
$n(z)$ is the electron density, and $N_D(z)$ is the density of positive
donor charges, which are assumed to be located far from the quantum
wells.
Integrating Eq.\ (\ref{eq:poisson}) twice, one obtains
\begin{equation}
V_H(z)=-\frac{4 \pi e^2}{\epsilon}\left(\int_{0}^{z} dz' (z-z') n(z') +
z \int_{-\infty}^0 dz' n(z') - \frac{N_s}{2} z \right), \label{eq:hartree}
\end{equation}
where $N_s$ is the electronic sheet density.
For the exchange-correlation potential, we use the parametrization due to 
Ceperley and Alder\cite{cep-ald} given by Eq.~(\ref{eq:LSDA-Vxc})
in Sec.~\ref{sec:LSDA} for both the spin-polarized and -unpolarized cases.

Once the subband energies $\varepsilon_n$ and wave functions $\phi_n(z)$ 
are obtained by solving Eq.\ (\ref{eq:kohn-sham}) numerically, the
$z$-dependent electron density is calculated from
\begin{equation}
n(z) =
  g_s \sum_{n {\bf k}} f(E_{n {\bf k}}) \,
  \left| \Psi_{n {\bf k}} ({\bf R}) \right|^2,
\label{eq:density1}
\end{equation}
where the factor $g_s$ accounts for the spin degeneracy and $f(E)$
= $1 / (e^{\beta E} + 1)$ is the Fermi-Dirac distribution function
with $\beta = 1 / T$ the inverse temperature ($k_B = 1$ throughout
this paper).
This density may be rewritten as
\begin{equation}
n(z) = \sum_n n_n \left| \phi_n (z) \right|^2
\label{eq:density2}
\end{equation}
with the subband occupancy $n_n$ given by
\begin{equation}
n_n = \frac{g_s}{A} \sum_{\bf k} f(E_{n {\bf k}}).
\end{equation}
The chemical potential is determined implicitly by the relation
\begin{equation}
N_s = \int dz \, n(z) = \sum_n n_n.
\label{eq:density3}
\end{equation}
At zero temperature, $f(E) \rightarrow \Theta(-E)$ and $n_n$ becomes
\begin{equation}
n_n = g_s N_0 (\varepsilon_F - \varepsilon_n) \,
  \Theta(\varepsilon_F - \varepsilon_n),
\label{eq:nn-zero}
\end{equation}
where $\Theta$ is the step function, $\varepsilon_F$ is the
Fermi energy, and $N_0 = m^* / 2 \pi \hbar^2$ is the two-dimensional,
single-spin density of states.

The LDA electronic structure for the spin-unpolarized case is thus 
obtained by solving Eq.\ (\ref{eq:kohn-sham}) together with
Eq.~(\ref{eq:hartree}) for $V_H[n(z)]$, Eq.~(\ref{eq:LSDA-Vxc}) for
$V_{XC}[n(z)]$, and Eq.~(\ref{eq:density2}) for $n(z)$ self-consistently.
The results of applying this procedure to a typical double-quantum-well
structure are shown in Fig.~\ref{figure1}.

From the wave functions and eigenenergies of the LDA calculation,
we can compute the collective modes of the confined electron gas
which are visible in inelastic light scattering experiments.
The relation between the cross section of resonant inelastic
light scattering by electronic excitations was obtained by Hamilton
and McWhorter\cite{ham-mcw} for bulk systems was adapted to the case of
semiconductor heterostructures by Katayama and Ando.\cite{kat-and}
These authors showed that the cross section for inelastic light scattering
by CDEs is proportional to the imaginary part of the {\em reducible}
electronic polarizability function $\tilde{\Pi}$ with the proportionality
factors depending on the details of the band structure of the host material.
These factors vanish for perpendicular polarizations of the incoming
and scattered light, and are maximized for parallel polarizations.
Thus, CDE spectra are measured in practice within the so-called 
{\em polarized} configuration, i.e., with parallel polarizations of the
two beams.
On the other hand, the scattering cross section due to spin-density
excitations is proportional to the imaginary part of the {\em irreducible}
electronic polarizability function $\Pi$, which is also called the
spin-polarizability function, and contains prefactors that are maximized
for perpendicular polarizations of the incident and scattered beams.
This is the usual geometry employed in measurements of SDE spectra, and is
referred to as the {\em depolarized} configuration.
Since we are interested in the properties of the electron gas confined in
the semiconductor structure, it is sufficient to calculate the
electronic response functions $\tilde{\Pi}$ and $\Pi$ and ignore the
band-structure-dependent factors. 

We compute these response functions within the TDLDA, which
is equivalent to calculating the irreducible polarizability function
including a static, $q$-independent vertex correction in the
ladder diagram approximation.\cite{mar-das}
Within this approximation, the integral equation for the irreducible
polarizability $\Pi(q,\omega)$ can be solved exactly and gives
\begin{equation}
\Pi(q,\omega)= \frac{\Pi^0(q,\omega)}{1+U_{XC} \, \Pi^0(q,\omega)},
                                                     \label{eq:pi-irred}
\end{equation}
where $\Pi^0(q,\omega)$ is the leading-order polarizability function
and $U_{XC}$ is the static, $q$-independent vertex function.
From Dyson's equation for the effective Coulomb interaction,\cite{fet-wal}
one obtains the reducible polarizability function
\begin{equation}
\tilde{\Pi}(q,\omega)= \frac{\Pi(q,\omega)}{1-U_H(q) \, \Pi(q,\omega)},
                                                     \label{eq:pi-red}
\end{equation}
where $U_H(q)$ is the Fourier transform of the bare Coulomb interaction.
Combining Eqs.\ (\ref{eq:pi-irred}) and (\ref{eq:pi-red}), we obtain
\begin{equation}
\tilde{\Pi}(q,\omega)= \frac{\Pi^0(q,\omega)}{1-(U_H(q)-U_{XC}) \,
                       \Pi^0(q,\omega)}.
\end{equation}

In a confined electron gas system, where the confinement discretizes
the single-particle energy levels, the collective excitations must
be calculated within a generalized dielectric function
formalism.\cite{sds-light-scatt}
In this context, the functions $U_{XC}$ and $U_H(q)$ are replaced
by matrices with indices labeling the different subbands.
Within TDLDA, we have
\begin{equation}
U_{ij,mn}^{XC} = - \int dz \, \int dz' \, \phi_i(z) \phi_j(z) \,
                   \frac{\partial V_{XC}}{\partial n}(z) \delta(z-z')
                   \, \phi_m(z') \phi_n(z')
                                                           \label{eq:UXC}
\end{equation}
and
\begin{equation}
U_{ij,mn}^H(q) = \frac{2 \pi e^2}{\epsilon q} \int dz dz' \, \phi_i(z)
                 \phi_j(z) \, e^{-q|z-z'|} \, \phi_m(z') \phi_n(z'),
                                                             \label{eq:UH}
\end{equation}
where $\epsilon$ is the background dielectric constant.

The reducible polarizability function $\tilde{\Pi}({\bf q},q_z,\omega)$,
whose imaginary part is proportional to the spectrum of the CDEs
and to the Raman intensity in the polarized configuration, is given
by\cite{jai-das}
\begin{equation}
\tilde{\Pi}({\bf q},q_z,\omega) = \int \, dz \int \, dz' e^{-iq_z (z-z')}
\, \tilde{\Pi}(z,z';{\bf q},\omega)
                                                     \label{eq:Pi-tilde}
\end{equation}
with
\begin{equation}
\tilde{\Pi}(z,z';{\bf q},\omega) = \sum_{i,j,k,l} \phi_i(z) \phi_j(z) \,
                    \tilde{\Pi}_{ij,kl}({\bf q},\omega) \, \phi_k(z')
                    \phi_l(z'),
\end{equation}
\begin{equation}
\tilde{\Pi}_{ij,kl}({\bf q},\omega) = \Pi_{ij}^0({\bf q},\omega)
                                      \, \delta_{ik} \, \delta_{jl} +
                                      \sum_{m,n}
                                      \Pi_{ij}^0({\bf q},\omega) \,
                                      U_{ij,mn}(q) \,
                                      \tilde{\Pi}_{mn,kl}({\bf q},\omega),
\end{equation}
\begin{equation}
U_{ij,mn}({\bf q}) = U_{ij,mn}^H({\bf q}) - U_{ij,mn}^{XC},
                                                             \label{eq:U}
\end{equation}
and
\begin{equation}
\Pi_{ij}^0({\bf q},\omega) = 2 \sum_{\bf k} \frac
             {f(E_j({\bf k+q})) - f(E_i({\bf k}))}
             {E_j({\bf k+q})) - E_i({\bf k})) - \hbar(\omega + i \gamma)}.
                                                              \label{eq:pi0}
\end{equation}
In these equations, subscripts are the subband indices,
${\bf q}$ and ${\bf k}$ are two-dimensional in-plane wave vectors, and
$\phi_j$ and $\varepsilon_j$ are the LDA-calculated subband wave functions
and energies.
In addition, $\Pi_{ij}^0$ is the leading-order polarizability function
for the transition $i \rightarrow j$,
$E_j({\bf k}) = \varepsilon_j + \frac{\hbar^2 k^2}{2m^*}$,
$f(E)$ is the Fermi factor, and $\gamma$ is a phenomenological inverse
scattering time; at $T = 0$, an analytic expression for
$\Pi_{ij}^0$ can be found in Ref.~\onlinecite{jai-das}.
We note that the random-phase approximation (RPA) is obtained
in the subband representation by removing the vertex correction
$U_{ij,mn}^{XC}$ in Eq.\ (\ref{eq:U}).

The imaginary part of the irreducible polarizability function $\Pi$ is
proportional to the SDE spectrum and to the Raman-scattering intensity
in the depolarized configuration.
In the subband representation, the calculation of $\Pi$ is analogous to
that of $\tilde{\Pi}$ [Eqs.\ (\ref{eq:UH})-(\ref{eq:pi0})]
with the following two modifications.
Since the irreducible polarizability does not include dynamic Coulomb
screening (spin-density excitations are unscreened by the spin-conserving
Coulomb interaction), we set $U_{ij,mn}^H=0$ in Eq.\ (\ref{eq:U}).
The second change concerns the vertex correction $U_{ij,mn}^{XC}$,
which for spin-density excitations is given in the TDLDA by 
\begin{equation}
U_{ij,mn}^{XC} = - \int dz \, \int dz' \, \phi_i(z) \phi_j(z) \,
                   \frac{\partial V_{XC}}{\partial m}(z) \delta(z-z')
                   \, \phi_m(z') \phi_n(z')
                                                           \label{eq:UXCm}
\end{equation}
instead of by Eq.\ (\ref{eq:UXC}).
In this equation, $m(z) = n_\uparrow(z)-n_\downarrow(z)$ is the local
spin density, $n_\uparrow$ and $n_\downarrow$ being the spin-up and
spin-down local densities, respectively.

The CDE and SDE energies are given by the poles of $\tilde{\Pi}$
and $\Pi$, respectively, which occur when the determinant
$|\Pi_{ij}^0 U_{ij,mn} - \delta_{ij,mn}|$ vanishes.
In the numerical work presented in this paper, we solve this equation
keeping all the subband levels.
If, as in the quantum well structures we consider, the lowest two
subbands are well separated in energy from the higher subbands, then
one can approximate this determinental equation in the limit of low
densities and temperatures by keeping only subbands 1 and 2, yielding
\begin{equation}
(\Pi_{12}^0 + \Pi_{21}^0) U_{12,12} = 1.   \label{eq:intersubband}
\end{equation}
For $q \rightarrow 0$, this condition gives the resonance energies
in the familiar form of Ando:\cite{ando76} 
\begin{equation}
\hbar^2 \tilde{\omega}_{21}^2 = \varepsilon_{21}^2 + 2 \varepsilon_{21}
                        U_{12,12} (N_1 - N_2),        \label{eq:ando76}
\end{equation}
where $\varepsilon_{21} \equiv \varepsilon_2-\varepsilon_1$ and 
$U_{12,12}=U_{12,12}^H(0)-U_{12,12}^{XC}$ 
in the case of the CDE, and $U_{12,12}=-U_{12,12}^{XC}$ for the SDEs.

\subsubsection{Polarized Electron Gas (LSDA)}
\label{sec:LSDA}

In this Subsection, we introduce a generalization of the
local-density approximation to density-functional theory discussed
in Sec.~\ref{sec:LDA-unpolarized} which allows for different populations
of the two spin orientations, i.e., a finite spin polarization.
This local-spin-density approximation (LSDA) formalism is also based
on the self-consistent solution of the Schr\"{o}dinger-like Kohn-Sham
equation, coupled with the Poisson equation and a local
exchange-correlation potential.
The main technical difference between LSDA and LDA is that the
effective exchange-correlation potential in LSDA depends on the local
spin polarization as well as the electron density.
Therefore, one has to solve two Kohn-Sham equations, which
contain spin-dependent effective potentials, for the two components
of the spinor wave function.
The LSDA was first formally justified by von Barth and Hedin\cite{von-hed}
and Pant and Rajagopal\cite{pan-raj} and is suitable for studying
ferromagnetic systems either with or without an external
magnetic field.\cite{dre-gro}

To put the LSDA approach in context, we briefly review the theoretical
and numerical evidence that the uniform electron gas in two and three
dimensions embedded in a uniform positive background (the jellium model)
undergoes a ferromagnetic transition at a certain critical density.
A simple theoretical estimate for the density at which a
ferromagnetic state will form may be obtained from Hartree-Fock theory,
which treats the electron-electron Coulomb interaction to first order.
For a uniform electron gas in three dimensions with $N_{+}$
spin-up and $N_{-}$ spin-down electrons, the ground-state energy
in this approximation can be written in terms of the total number
of particles $N = N_{+} + N_{-}$ and the magnetization
$m = (N_{+} - N_{-} ) / N$ as\cite{fet-wal}
\begin{eqnarray}
E_{HF}^{3D}
&=& \frac{Ne^2}{2a_0}
    \frac{3}{10} \left(\frac{9\pi}{2}\right)^{2/3} \frac{1}{r_s^2}
    \left[ \left(\frac{1+m}{2}\right)^{5/3} + 
           \left(\frac{1-m}{2}\right)^{5/3}\right]        \nonumber \\
& & -\frac{Ne^2}{2a_0}  
      \frac{3}{4\pi} \left(\frac{9\pi}{2}\right)^{2/3} \frac{1}{r_s}
      \left[ \left(\frac{1+m}{2}\right)^{4/3} + 
             \left(\frac{1-m}{2}\right)^{4/3}\right],
\label{eq:hartreefockenergy}
\end{eqnarray}
where $a_0=\hbar^2 / m^* e^2$ is the Bohr radius and
$r_s = (3V/4\pi N)^{1/3}/a_0$ parameterizes the density.
The first term in this expression is the kinetic energy, which prefers
the paramagnetic state, while the second term is the exchange energy,
which prefers to polarize the spins.
At densities satisfying $r_s > 5.45$, the exchange energy dominates,
and the ferromagnetic state is stable.
In two dimensions, a similar analysis yields\cite{Rajagopal}
\begin{eqnarray}
E_{HF}^{2D} &=& \frac{Ne^2}{2a_0}
      \left\{ \frac{1+m^2}{r_s^2} - \frac{4 \sqrt{2}}{3 \pi r_s}
        \left[(1+m)^{3/2} + (1-m)^{3/2}\right] \right\},
\end{eqnarray}
where now $r_s \equiv (A/\pi N)^{1/2} / a_0$.
In this case, the condition for a polarized ground-state,
$E_{HF}^{2D}(r_s,m=1) < E_{HF}^{2D}(r_s,m=0)$, is satisfied if $r_s > 2.01$.
Thus, within the Hartree-Fock approximation, the spin-polarized state
occurs at higher density (lower $r_s$) than in three dimensions.

Hartree-Fock theory neglects contributions to the energy beyond the
exchange term and is therefore expected to overestimate the density
at which the ferromagnetic transition occurs.
Including these so-called correlation terms can only be done approximately,
however.
Currently, the most accurate method for performing these calculations
are numerically intensive Monte Carlo techniques.
Ceperley\cite{cep-ald} calculated the ground-state energy of an 
electron gas in two and three dimensions employing variational
Monte Carlo (VMC).
He found that in both two and three dimensions there is an intermediate
density regime where a fully polarized state has the lowest energy
compared to the unpolarized quantum liquid and the Wigner crystal.
In 3D, the polarized phase is stable for $26<r_s<67$, and, in 2D,
for $13<r_s<33$.
Additional results seem to indicate that in 3D there is a transition to
a partially polarized liquid at $r_s \approx 20$ and to a fully
polarized phase at $r_s \approx 50$.\cite{cep-unp}
Eleven years later, Tanatar and Ceperley\cite{tan-cep} recalculated 
the ground-state properties of the electron gas in two dimensions 
employing the VMC technique and the more accurate fixed-node 
Green's-function Monte Carlo (GFMC) technique.
The VMC technique predicted again a transition from the unpolarized
to the polarized liquid at $r_s$ between 10 and 20, consistent
with Ceperley's results.
The more accurate GFMC technique predicted a transition from
the unpolarized liquid to the Wigner crystal at $r_s=37$, without 
an intermediate polarized phase.
However, the authors point out that, near the transition,
the polarized phase has an energy very close to the energy of the 
other phases and that, due to finite size effects and errors associated
with their approximation method, their conclusion should not be 
taken as definite.
This leaves open the possibility of a stable, fully-polarized phase
in the two-dimensional electron gas.

The LSDA falls somewhere between elementary Hartree-Fock theory and
Monte Carlo calculations in terms of the quantitative accuracy of
its predictions.
Its main strength, and the reason we use this technique here, is that
it can describe the inhomogeneous electron gas which exists inside
quantum well structures and can therefore provide estimates of the
critical density of the ferromagnetic transitions in these structures
that could guide future experiments.
In addition, the LSDA is a direct extension of the LDA and so allows
a comparison between the two calculations which will be important
in ruling out ferromagnetism as the source of the spin-density-excitation
softening which appears in our TDLDA calculations
[Cf. Sec.~\ref{sec:instability}].
We note that a similar problem to that of the spin polarization of
the ground-state in quantum wells is the problem of ``valley condensation''
in Si-SiO$_2$ systems, where, instead of spin states, the electrons can
occupy different valleys of the Brillouin zone. 
This problem has been studied in the past with techniques similar to the
ones employed here.\cite{das-vin:val-deg}
In addition, the LSDA has been employed to study spin effects in wide
parabolic quantum wells in the presence of a perpendicular magnetic
field.\cite{hem-mas-kwi}

Our computations in the LSDA proceed as follows.
After factorizing the complete single-electron wave function as 
was done in Eq.\ (\ref{eq:factorize}), we write down the $z$-dependent
Kohn-Sham equation:
\begin{equation}
\left(-\frac{\hbar^2}{2m^*} \frac{d^2}{dz^2} + V_C(z) +V_H(z) + 
V_{XC}^{\sigma}(z) \right) \phi_n^{\sigma}(z) =
\varepsilon_n^{\sigma} \phi_n^{\sigma}(z),
                                                  \label{eq:ferro-kohn-sham}
\end{equation}
where $n$ is the subband index and $\sigma$ denotes
the spin orientation, which can be up (+) or down (-).
We assume that no more than two subbands are populated and let $n = 1$
denote the lower-energy symmetric level (S) and $n = 2$ the
higher-energy antisymmetric level (AS).
We therefore need to consider four wave functions ($\phi_1^+$,
$\phi_1^-$, $\phi_2^+$, $\phi_2^-$) and their corresponding energies
in Eq.~(\ref{eq:ferro-kohn-sham}).

This equation also contains the exchange-correlation potential,
which in the LSDA formalism depends on both the density $n(z)$ and the
spin polarization $m(z)$, which is defined as
\begin{equation}
m(z) \equiv \frac{n_1^+(z)+n_2^+(z)-n_1^-(z)-n_2^-(z)}{n(z)}.
\end{equation}
In our calculations, we use the parametrization of the exchange-correlation
energy for the uniform 3D electron gas obtained by Ceperley 
and Alder:\cite{cep-ald}
\begin{equation}
\epsilon_{XC}^i = \frac{c^i}{r_s} +
\frac{\gamma^i}{1+\beta_1^i \sqrt{r_s}+\beta_2^i \, r_s},
\end{equation}
where $i=U$ (unpolarized, $m=0$) or $i=P$ (polarized, $m=1$).
The exchange-correlation contribution to the chemical potential is
\begin{eqnarray}
V_{XC}^i &=& (1 - \frac{r_s}{3} \frac{d}{dr_s}) \, \epsilon_{XC}^i
                                                                  \nonumber \\
&=& \frac{d^i}{r_s} + \gamma^i \,
\frac{1 + \frac{7}{6} \, \beta_1^i \sqrt{r_s} + \frac{4}{3} \, \beta_2^i \, r_s}
     {(1 + \beta_1^i \sqrt{r_s} + \beta_2^i \, r_s)^2}      \label{eq:cep-ald}
\end{eqnarray}
The parameters in the previous expressions, as obtained by
Ceperley and Alder,\cite{cep-ald} are:
$c^U = -0.9163$,
$c^P = -1.1540$,
$d^U = -1.2218$,
$d^P = -1.5393$,
$\gamma^U = -0.1423$,
$\gamma^P = -0.0843$,
$\beta_1^U = 1.0529$,
$\beta_1^P = 1.3981$,
$\beta_2^U = 0.3334$, and
$\beta_2^P = 0.2611$.
For intermediate polarizations ($0<m<1$), we use an interpolation
formula proposed by von Barth and Hedin,\cite{von-hed} in which the
correlation energy has the same polarization dependence as the
exchange energy:
\begin{equation}
\epsilon_{XC}(r_s,m) = \epsilon_{XC}^U(r_s) + f(m)
  \left(\epsilon_{XC}^P(r_s) - \epsilon_{XC}^U(r_s) \right)
\end{equation}
and
\begin{eqnarray}
V_{XC}^{\sigma}(r_s,m) &=& V_{XC}^U(r_s) + f(m)
  \left(V_{XC}^P(r_s) - V_{XC}^U(r_s)\right) +             \nonumber \\
  && \left(\epsilon_{XC}^P(r_s) - \epsilon_{XC}^U(r_s)\right)
  \left(\mbox{sgn}(\sigma) - m \right) \frac{df}{dm} , \label{eq:LSDA-Vxc}
\end{eqnarray}
where
\begin{equation}
f(m) = \frac{(1+m)^{4/3} + (1-m)^{4/3} -2}{2^{4/3} - 2}.
\end{equation}
The Hartree potential $V_H(z)$ is calculated as in the unpolarized
formalism described in Sec.~\ref{sec:LDA};
it satisfies Poisson equation, Eq.\ (\ref{eq:poisson}), and is given
by Eq.\ (\ref{eq:hartree}).

To complete the specification of the problem, we note that
the density associated with each subband and spin orientation
is given by
\begin{equation}
n_n^{\sigma}(z) = n_n^{\sigma} \, |\phi_n^{\sigma}(z)|^2,
\end{equation}
where $n_n^{\sigma}$ is the occupancy of each level, which at zero
temperature is given by
\begin{equation}
n_n^{\sigma} = N_0 \, (\varepsilon_F - \varepsilon_n^\sigma) \,
  \Theta(\varepsilon_F - \varepsilon_n^\sigma).
\end{equation}
The total electron density may be written
\begin{equation}
n(z) = \sum_{n \, \sigma} n_n^\sigma(z),
\end{equation}
which implies that the Fermi level $E_F$ is implicitly determined by
the condition
\begin{equation}
N_s = \int_{-\infty}^{\infty} dz \, n(z) = \sum_{n \, \sigma} n_n^\sigma
\end{equation}
[cf. Eqs.~(\ref{eq:density1})-(\ref{eq:nn-zero})].
The self-consistent solution of these equations proceeds exactly as in
the LDA case with the difference that now, in each iteration, one has
to solve two Kohn-Sham equations, Eq.\ (\ref{eq:ferro-kohn-sham}),
for the two spinor components of the wave functions, $\phi_n^\sigma$.

\subsection{Self-Consistent Hartree-Fock Theory}
\label{sec:SCHF}

While conventional density-functional theory is a fairly accurate
method for determining the properties of semiconductor heterostructures
in their paramagnetic and ferromagnetic phases, it is unable to address
more complicated magnetic ordering such as antiferromagnetism.
To study such phases, it is useful to return to a model Hamiltonian
of the electronic system and search for the existence of
broken-symmetry states within a mean-field theory.
In some cases, such as superconductivity, the mean-field theory
gives a quantitative account of these phases.\cite{Schrieffer}
More commonly, however, it sacrifices quantitative accuracy in favor
of qualitative insight.
This insight manifests itself not only in a physical intuition about
the nature of the new ground state but also in the ability to study
the distinctive features of the broken-symmetry phase, which may
serve as a guide for interpreting experimental data in these systems.
We adopt this point of view in what follows.

To that end, consider a three-dimensional electron gas interacting
through a potential $V ({\bf R}-{\bf R'})$ and confined along
the $z$-direction by a potential $V_{CONF} (z)$.
A confining potential of this type is shown in Fig.~\ref{figure1} for a
double quantum well, but the precise shape is unimportant for
the development of the formalism.
Given a particular $V_{CONF} (z)$, one can construct its eigenfunctions
$\xi_n (z)$ and eigenenergies $\epsilon_n$, $n$ = 1, 2, 3...,
by solving the time-independent Schr\"{o}dinger equation
\begin{equation}
\left[ -\frac{\hbar^2}{2m^*} \frac{d^2}{dz^2} + V_{CONF} (z) \right]
  \xi_n (z) = \epsilon_n \xi_n (z).
\label{eq:xidef}
\end{equation}
In terms of these eigenfunctions, the quasiparticle annihilation operator
$\psi_{\sigma} ({\bf R})$ can be written as
\begin{equation}
\psi_{\sigma} ({\bf R}) =
  \frac{1}{\sqrt{A}} \sum_{n{\bf k}} e^{i{\bf k \cdot r}} \xi_n (z)
  c_{n{\bf k}\sigma},
\label{eq:basis}
\end{equation}
where ${\bf R} = ({\bf r},z) = (x,y,z)$, ${\bf k} = (k_x, k_y)$,
$A$ is the transverse area of the sample, and $c_{n{\bf k}\sigma}$
annihilates a quasiparticle in subband $n$, of transverse wave vector
${\bf k}$, and with spin projection $\sigma$ (these conventions will
be used throughout this paper).
Note that, unlike the density-functional approach, the total confining
potential $V_{CONF} (z)$ is specified at the outset and is not
determined self-consistently.

Defining a composite subband and spin index $a = (n_a, \sigma_a)$
with summation over repeated indices implied, the Hamiltonian
may be written in the basis defined by Eq.~(\ref{eq:basis}) as
\begin{eqnarray}
H &=& H_0 + H_{\rm int} \nonumber \\
&=& \sum_{\bf k} \, \epsilon_{a{\bf k}}^{~} \,
  c^{\dag}_{a {\bf k}} c^{~}_{a {\bf k}}
+ \frac{1}{2 A} \, \sum_{{\bf k} {\bf k'} {\bf q}}
  \, V_{ad,bc}^{~} ({\bf q}) \,
  c^{\dag}_{a {\bf k + q}} c^{\dag}_{b {\bf k' - q}}
  c^{ }_{c {\bf k'}} c^{ }_{d {\bf k}} .
\label{eq:H}
\end{eqnarray}
In this expression, the quasiparticle energy
\begin{equation}
\epsilon_{a{\bf k}} = \epsilon_n + \frac{\hbar^2 k^2}{2m^*} - \mu
\label{eq:ea}
\end{equation}
is measured with respect to the chemical potential $\mu$,
and the matrix elements of the interaction are
\begin{eqnarray}
V_{ab,cd} ({\bf q}) &=&
  \delta_{\sigma_a \sigma_b} \delta_{\sigma_c \sigma_d} \,
  \int \frac{d{\bf R} \, d{\bf R'}}{A} \,
  e^{i{\bf q \cdot} ({\bf r - r'})} \,
  \xi_{n_a}^* (z) \xi_{n_b}^{~} (z) \, V ({\bf R} - {\bf R'}) \,
  \xi_{n_c}^{*} (z') \xi_{n_d}^{~} (z') .
\label{eq:V}
\end{eqnarray}

Our goal is to solve this model within mean-field theory allowing
for the possibility of broken-symmetry phases.
For reasons that will be discussed in Sec.~\ref{sec:AF}, we shall
restrict attention to those ground states which are translationally
invariant transverse to the layering direction, but we shall allow for
off-diagonal order in both the subband and spin indices.
This assumption excludes from the outset the study of intra-well charge-
or spin-density waves and Wigner crystallization, and it is also
implicit in the density-functional calculations discussed in the
preceding Subsection.
One could modify our treatment to include such phases, but the present
model is sufficient for the purposes of exploring the effects of the
inter-well degrees of freedom.
The assumption of translational invariance implies conservation of
the transverse wave vector, and so the quasiparticle propagator
can be written
\begin{equation}
G_{ab} (k_n) = - \int_0^{\beta} d\tau \, e^{i\omega_n \tau} \,
  \left< T_\tau \left[ c_{a{\bf k}}^{~} (\tau) c_{b{\bf k}}^{\dag} (0)
  \right] \right>,
\label{eq:G}
\end{equation}
where $k_n = ({\bf k},i\omega_n)$, $\beta = 1 / T$
($\hbar = k_B = 1$ throughout this paper), and the rest of the
notation is standard.\cite{Mahan}

The mean-field theory for our model is constructed by using
this propagator to compute the electronic self-energy in the
self-consistent Hartree-Fock approximation.
This approximation corresponds to expanding the self-energy to
one-loop order in the interaction and is shown diagrammatically
in Fig.~\ref{fig:diagrams}(a).
The resulting self-energy is
\begin{equation}
\Sigma_{ab} ({\bf k}) =
  \frac{T}{A} \sum_{k^{\prime}_m} e^{-i \omega_m 0-} \,
  \left[ V_{ab,dc} ({\bf 0}) - V_{ac,db} ({\bf k - k^{\prime}}) \right] \,
  G_{cd} (k^{\prime}_m) .
\label{eq:Sigma}
\end{equation}
The self-consistency of this approximation arises because the
propagators used in Eq.~(\ref{eq:Sigma}) are dressed by the
same self energy according to the Dyson equation
\begin{equation}
\left[ \left( i\omega_n - \epsilon_{a{\bf k}} \right) \delta_{ab}
  - \Sigma_{ab} ({\bf k}) \right] \, G_{bc} (k_n) = \delta_{ac}.
\label{eq:Dyson}
\end{equation}
To completely specify the system of equations, the chemical potential
is determined from the band-filling constraint
\begin{equation}
N_s = \frac{T}{A} \sum_{k_m} e^{-i \omega_m 0-} \, G_{aa} (k_m) .
\label{eq:filling}
\end{equation}

An alternative form of these equations, which will turn out to be
convenient for future work, is obtained by inverting the Dyson
equation for the interacting propagator $G_{ab} (k_n)$
[Eq.~(\ref{eq:Dyson})] by making an appropriate choice of the basis.
From Eqs. (\ref{eq:V}) and (\ref{eq:Sigma}), the matrix
$\epsilon_{a{\bf k}} \delta_{ab} + \Sigma_{ab} ({\bf k})$
is Hermitian and so possesses a complete and orthonormal set of
eigenfunctions $\varphi^c_a ({\bf k})$.
These eigenfunctions satisfy the eigenvalue equation
\begin{equation}
\left[ \epsilon_{a{\bf k}} \delta_{ab} + \Sigma_{ab} ({\bf k}) \right] \,
  \varphi^c_b ({\bf k}) = E^c ({\bf k}) \, \varphi^c_a ({\bf k})
\label{eq:eigen_eqn}
\end{equation}
(no sum on $c$) and the orthonormality relations
\begin{equation}
\varphi^c_a ({\bf k}) \varphi^{*c}_b ({\bf k}) = \delta_{ab}
\end{equation}
and
\begin{equation}
\varphi^a_c ({\bf k}) \varphi^{*b}_c ({\bf k}) = \delta^{ab}.
\label{eq:ON}
\end{equation}
Eq.~(\ref{eq:Dyson}) is diagonal in the basis of these eigenfunctions
by construction, so we may invert the equation to obtain
\begin{equation}
G_{ab} (k_n) =
  \frac{\varphi^c_a ({\bf k}) \, \varphi^{c*}_b ({\bf k})}
  {i\omega_n - E^c ({\bf k})}.
\label{eq:Gspec}
\end{equation}
Substituting this relation back into the equation for the self-energy
[Eq.~(\ref{eq:Sigma})] and performing the sum over Matsubara frequencies
yields
\begin{equation}
\Sigma_{ab} ({\bf k}) =
  \frac{1}{A} \sum_{\bf k^{\prime}} \,
  \left[ V_{ab,dc} ({\bf 0}) - V_{ac,db} ({\bf k - k^{\prime}}) \right] \,
  \varphi^e_c ({\bf k^{\prime}}) \varphi^{*e}_d ({\bf k^{\prime}}) \,
  f(E^e ({\bf k^{\prime}})),
\label{eq:Sigma2}
\end{equation}
where $f(x) = 1 / (e^{\beta x} + 1)$ is the Fermi function.
The band filling constraint Eq.~(\ref{eq:filling}) may similarly be
written
\begin{equation}
N_s = \frac{1}{A} \sum_{c{\bf k}} f(E^c ({\bf k}))
  \equiv \sum_c n^c.
\label{eq:n}
\end{equation}
Eqs.~(\ref{eq:eigen_eqn})-(\ref{eq:n}) are the equations we will
ultimately solve for a simple model interaction in Sec.~\ref{sec:model}.

The eigenfunctions $\varphi^c_a ({\bf k})$ and eigenenergies $E^c ({\bf k})$
of the operator $\epsilon_{a{\bf k}} \delta_{ab} + \Sigma_{ab} ({\bf k})$
are the wave functions and energies of the quasiparticles of the
interacting system within self-consistent Hartree-Fock theory.
An alternative way of saying the same thing is that we have performed
a mean-field decomposition of the Hamiltonian and diagonalized
the result with a Bogoliubov transformation in the particle-particle
channel.
The annihilation operators for the interacting quasiparticles
$\gamma^c_{\bf k}$ are therefore obtained from the bare operators
via $\gamma^c_{\bf k} = \varphi^c_a ({\bf k}) c_{a{\bf k}}$.

In addition to the physical insight afforded by rewriting the
self-energy equations in terms of these functions, the calculation
of the energy, entropy, and specific heat of the system becomes
straightforward.
The energy is the expectation value of the Hamiltonian
[Eq.~(\ref{eq:H})] with the energy shift due to the chemical
potential removed:
\begin{eqnarray}
E &=& \left< H + \mu N \right> \nonumber \\
  &=& \left< H_0 \right> + \left< H_{\rm int} \right> + \mu N_s.
\label{eq:Etot}
\end{eqnarray}
Within our self-consistent Hartree-Fock theory, the contribution
to the energy from the interaction term in the Hamiltonian is
shown graphically by the diagrams in Fig.~\ref{fig:diagrams}(b).
These diagrams lead to the result
\begin{equation}
\left< H_{\rm int} \right> =
  \frac{T}{2} \sum_{k_n} e^{-i\omega_n 0-} \, G_{ba} (k_n) \,
  \Sigma_{ab} ({\bf k}),
\label{eq:Eint}
\end{equation}
which allows us to write Eq.~(\ref{eq:Etot}) as
\begin{equation}
E = \frac{T}{2} \sum_{k_n} e^{-i\omega_n 0-} \, \left[
  2 \epsilon_{a{\bf k}}^{~} \delta_{ab} + \Sigma_{ab} ({\bf k})
  \right] G_{ba} (k_n) + \mu N_s.
\label{eq:Eg}
\end{equation}
Substituting Eq.~(\ref{eq:Gspec}) into this equation, using the
eigenequation Eq.~(\ref{eq:eigen_eqn}) and orthonormality relation
Eq.~(\ref{eq:ON}), and performing the Matsubara sum, the total
energy becomes
\begin{equation}
E = \frac{1}{2} \sum_{\bf k} \left[
  \left| \phi^{c}_a ({\bf k}) \right|^2 \epsilon_{a{\bf k}}
  + E^c ({\bf k}) \right] f(E^c({\bf k})) + \mu N_s.
\label{eq:Ephi}
\end{equation}
The entropy in the interacting basis is simply the standard
free-fermion result
\begin{equation}
S = -\sum_{\bf k} \left\{ f(E^c ({\bf k})) \ln f(E^c ({\bf k}))
  + \left[ 1 - f(E^c ({\bf k})) \right]
  \ln \left[ 1 - f(E^c ({\bf k})) \right] \right\},
\label{eq:S}
\end{equation}
and the specific heat is obtained directly from this equation:
\begin{eqnarray}
C_V &=& T \frac{\partial S}{\partial T} \nonumber \\
  &=& \sum_{\bf k}
  \left( -\frac{ \partial f (E^c ({\bf k}) ) }{\partial E^c ({\bf k}) } \right)
  \, \beta E^c({\bf k}) \, \frac{d (\beta E^c ({\bf k}))}{d \beta}.
\label{eq:Cv}
\end{eqnarray}

In order to get a full picture of the interacting system, one must
go beyond the single-particle properties and thermodynamic functions
and examine the response of the system to external perturbation.
As discussed in Sec.~\ref{sec:LDA-unpolarized}, resonant inelastic
light scattering has proven to be a powerful tool for studying the
charge- and spin-density excitations in semiconductor heterostructures.
Consequently, we will focus on the generalized density response function
and the resulting collective excitations which can be observed in
these experiments.
As in Sec.~\ref{sec:LDA-unpolarized}, we will not compute the form
factors necessary to connect the polarizability to the inelastic light
scattering cross section;
we merely note that the collective modes we will discuss are
detectable in these experiments with a particular (and structure-dependent)
arrangement of scattering angles, polarizations, etc.

We begin by defining a generalized density operator
\begin{equation}
\rho^{\mu} ({\bf R}) =
  \psi_{\sigma_1}^{\dag} ({\bf R}) \sigma_{\sigma_1 \sigma_2}^{\mu}
  \psi_{\sigma_2} ({\bf R}),
\label{eq:rho_r}
\end{equation}
where $\sigma^{\mu}$ = $\{ \sigma^0, \sigma^1, \sigma^2, \sigma^3 \}$
= $\{ 1, \sigma^x, \sigma^y, \sigma^z \}$.
Note that the number density $n ({\bf R}) = \rho^0 ({\bf R})$ and
the spin density $s^i ({\bf R}) = (\hbar / 2) \, \rho^i ({\bf R})$,
$i$ = 1, 2, or 3.
Suppose we can couple to this density through an external force
$F^{\mu}_{\rm ext} ({\bf R}, t)$ which adds a term
\begin{equation}
H_{\rm ext} (t) =
  - \int d{\bf R} \, \rho^{\mu} ({\bf R}, t) F^{\mu}_{\rm ext} ({\bf R}, t)
\end{equation}
to the Hamiltonian (a sum on $\mu$ is implied).
The linear response of the generalized density to this perturbation
is then given by\cite{Forster}
\begin{equation}
\left< \delta \rho^{\mu} ({\bf R}, t) \right> =
  i \int d{\bf R'} dt' \, \theta(t') \,
  \left< \left[ \rho^{\mu} ({\bf R}, t), \rho^{\nu} ({\bf R'}, t')
  \right] \right> \, F^{\nu}_{\rm ext} ({\bf R'}, t'),
\label{eq:response}
\end{equation}
where the angle brackets denotes the thermodynamic average in the
absence of $H_{\rm ext} (t)$.

As mentioned above, we assume that the interacting system is
translationally invariant in the transverse direction, and it
is also time-translation invariant.
Thus, we may introduce partial Fourier transforms
\begin{equation}
\rho^{\mu} ({\bf R}, t) =
  \frac{1}{A} \sum_{\bf q} \, \int \frac{d\omega}{2\pi} \,
  e^{i({\bf q \cdot r} - \omega t)} \, \rho^{\mu} ({\bf r}, z, t)
\label{eq:rho_q}
\end{equation}
and write Eq.~(\ref{eq:response}) as
\begin{equation}
\left< \delta \rho^{\mu} ({\bf q}, z, \omega) \right> =
  i \int dz' dt' e^{i\omega t'} \, \theta(t') \,
  \left< \left[ \rho^{\mu} ({\bf q}, z, t), \,\rho^{\nu} ({\bf -q}, z', t')
  \right] \right> \, F^{\nu}_{\rm ext} ({\bf q}, z', \omega),
\label{eq:response_z}
\end{equation}
From Eqs.~(\ref{eq:basis}) and (\ref{eq:rho_r}), we can rewrite
the Fourier components of the density operator as
\begin{equation}
\rho^{\mu} ({\bf q}, z) =
  \xi_{n_a}^{*} \xi_{n_b}^{~} \, \sigma_{\sigma_a \sigma_b}^{\mu}
  \, \rho_{ab} ({\bf q})
\label{eq:rho_a}
\end{equation}
with
\begin{equation}
\rho_{ab} ({\bf q}) =
  \sum_{\bf k} c_{a{\bf k}}^{\dag} c_{b{\bf k+q}}^{~}.
\label{eq:rho_c}
\end{equation}
Using this relation, we obtain the final form for the generalized
density response:
\begin{equation}
\left< \delta \rho^{\mu}_{ab} ({\bf q}, \omega) \right> =
  - \Pi_{ab,cd} ({\bf q}, \omega) \, F^{\rm ext}_{cd} ({\bf q}, \omega)
\end{equation}
where
\begin{eqnarray}
\Pi_{ab,cd} ({\bf q}, \omega) &=&
  -i \int dt \, e^{i\omega t} \ \left < \left[
  \rho_{ab}^{~} ({\bf q}, t), \,\rho_{cd}^{\dag} ({\bf q}, 0)
  \right] \right> , \label{eq:response_a} \\
\left< \delta \rho^{\mu} ({\bf q}, z, \omega) \right> &=&
  \xi_{n_a}^{*} (z) \, \xi_{n_b}^{~} (z) \, \sigma_{\sigma_a, \sigma_b}^{\mu}
  \, \left< \delta \rho_{ab} ({\bf q}, \omega) \right> ,
\end{eqnarray}
and
\begin{equation}
F^{\rm ext}_{cd} ({\bf q}, \omega) =
  \int dz' \, \xi_{n_c}^{~} (z') \, \xi_{n_d}^{*} (z') \,
  \sigma_{\sigma_c \sigma_d}^{\nu} \,
  F^{\nu}_{\rm ext} ({\bf q}, z', \omega) .
\label{eq:F_a}
\end{equation}

In order to obtain the subband- and spin-resolved
polarizability [Eq.~(\ref{eq:response_a})], we compute the
polarizability in Matsubara frequencies $i\nu_n = 2\pi n T$,
\begin{equation}
\Pi_{ab,cd} (q_n) =
  - \int_0^{\beta} d\tau e^{i\omega_n \tau} \,
  \left< T_\tau \left[
  \rho_{ab} ({\bf q},\tau) \rho_{cd}^{\dag} ({\bf q},0)
  \right] \right>
\end{equation}
($q_n = ({\bf q}, i\nu_n)$),
and analytically continue to real frequencies by the conventional
substitution $i\nu_n \rightarrow \Omega + i\delta$ (cf.
Ref.~\onlinecite{Mahan}).
This polarizability is calculated within a conserving
approximation\cite{Baym} using the diagrams shown in
Figs.~\ref{fig:diagrams}(c) and \ref{fig:diagrams}(d).
These diagrams yield the expressions
\begin{equation}
\Pi_{ab,cd} (q_n) =
  \frac{T}{A} \sum_{k_m} G_{ea} (k_m) \, G_{bf} (k_m + q_n) \,
  \gamma_{ef,cd} (k_m, k_m + q_n)
\label{eq:Pi}
\end{equation}
for the polarizability and 
\begin{eqnarray}
\gamma_{ab,cd} (k_m, k_m + q_n) &=& \delta_{ac} \delta_{bd}
  - \frac{T}{A} \sum_{k_{l}^{\prime}} \,
  \left[ V_{bf,ea} ({\bf k} - {\bf k^{\prime}})
  - V_{ba,ef} ({\bf q}) \right] \nonumber \\
&\times& G_{ge} (k_{l}^{\prime}) \, G_{fh} (k_{l}^{\prime} + q_n)
  \gamma_{gh,cd} (k_{l}^{\prime}, k_{l}^{\prime} + q_n)
\label{eq:gamma}
\end{eqnarray}
for the vertex function.
Contrary to the usual convention, we include the RPA screening
diagrams in the vertex function and thus do not distinguish between
reducible and irreducible polarizabilities.
The reason is that, in the broken-symmetry phases we will examine,
the propagators are no longer diagonal in spin, so these screening
terms enter into both the spin- and charge-density response functions,
reducing the usefulness of the division between reducible and
irreducible polarizabilities.
Note that even these approximate equations for the polarizability
must be solved numerically for a general interaction.
If the interaction is wave-vector-independent, on the other hand,
the vertex equation can be solved simply (cf. Sec.~\ref{sec:AF}).

The interpretation of the polarizability as a response function
can be used to visualize the real-space density fluctuations
it describes.
If one perturbs the system with the external force
$F^{\mu}_{\rm ext} ({\bf R}, t)$ $\propto$
$e^{i({\bf q \cdot r} - \omega t)}$,
Eqs.~(\ref{eq:response})-(\ref{eq:F_a}) imply that the response
in real space will be
\begin{equation}
\left< \delta \rho^{\mu} ({\bf R}, t) \right> \propto
  e^{i({\bf q \cdot r} - \omega t)} \, \xi_{n_a}^{*} (z) \, \xi_{n_b}^{~} (z) \,
  \sigma_{\sigma_a \sigma_b}^{\mu} \,
  \Pi_{ab,cd}^{~} ({\bf q}, \omega) f_{cd}^{\rm ext} ,
\label{eq:polarization}
\end{equation}
where $f_{cd}^{\rm ext}$ is related to how this force perturbs the
subband and spin indices.
For example, to examine the spin-density response between the lowest
two subbands,
\begin{equation}
f_{cd}^{\rm ext} =
  \left[ \delta_{n_c 1} \delta_{n_d 2} + \delta_{n_c 2} \delta_{n_d 1}
  \right] \, \delta_{\sigma_c \uparrow} \delta_{\sigma_d \downarrow}.
\label{eq:fext}
\end{equation}
This procedure works for general wave vector and frequency, but has
special significance when these quantities correspond to a collective
mode of the system, in which case the response gives the polarization
of the collective mode.
Note that, since the response function diverges at a collective
excitation, in practice one applies this formula by adding a small
imaginary part to the frequency to control this divergence.
As with the equations for the polarizability, this numerically
intensive approach for obtaining the collective mode polarization
is required for the broken-symmetry phases where the off-diagonal
terms make an analytic calculation difficult.

\section{Intersubband Instabilities}
\label{sec:instability}

In this Section, we shall study the instabilities of an electron gas
confined to coupled double-quantum-well (DQW) and wide single-quantum-well
structures by examining their intersubband collective excitations.
We begin by applying the formalism described in Sec.~\ref{sec:LDA}
to the typical GaAs / $\rm Al_x Ga_{1-x} As$ DQW structure shown in
Fig.~\ref{figure1}; similar samples were used in the experimental study
of Ref.~\onlinecite{decca}.
Although our TDLDA calculation includes all the subband energies
shown in the figure, we are only concerned with intersubband transitions
between the lowest two subbands (subbands 1 and 2), whose typical separation
$\Delta_{SAS} \equiv \varepsilon_2-\varepsilon_1 \approx$ 1~meV.
The exact values of $\Delta_{SAS}$ are plotted in Fig.~\ref{figure2},
along with the calculated long-wavelength energies of the intersubband
collective modes as a function of the sheet density.

Two distinctive features of these collective modes are seen in
Fig.~\ref{figure2}.
The first feature is that the CDE energy becomes less than $\Delta_{SAS}$
around $N_s \approx 0.2 \times10^{11}$ cm$^{-2}$.
This behavior, originally predicted in Ref.~\onlinecite{mar-das} for
single quantum wells, has recently been verified
experimentally\cite{ern-gon-sya} and will not be discussed further.
The second -- and for our purposes more important -- feature is that
the intersubband SDE energy goes to zero below a critical density
$N_C \approx 0.7 \times 10^{11}$ cm$^{-2}$ and becomes finite again
below $0.1\times10^{11}$ cm$^{-2}$.
This softening of the intersubband SDE mode indicates that the normal
system with a ``metallic'' Fermi surface is unstable at or below $N_C$,
since it can spontaneously create spin-reversed intersubband electron-hole
pairs (``triplet excitons'') at no cost in energy.
We conclude that there is a phase transition in this DQW at the critical
sheet density $N_C$ from the normal 2D electron liquid to a triplet
intersubband exciton liquid and that the system re-enters the normal 
phase at a lower density.
This electronic phase transition is due exclusively to exchange-correlation
effects which make the vertex correction or excitonic shift larger than
$\Delta_{SAS}$ [Eq.~(\ref{eq:ando76})] and thus cause the SDE to collapse.
The re-entrance of the normal phase at lower density is simply explained
by the fact that the vertex correction vanishes as $N_s \rightarrow 0$
according to Eq.~(\ref{eq:ando76}).
We note that the transition to a Wigner crystal occurs at much 
lower $N_s$ values than those considered in this paper and so does not
account for the SDE collapse.\cite{nei,tan-cep}

Additional evidence in support of the excitonic transition comes from
the density dependence of the mean-field vertex correction 
$\mid U_{XC} \chi_{12}^0 \mid$, which is plotted in Fig.~\ref{figure3}(a).
This vertex correction consists of the spin-polarized,
exchange-correlation-induced vertex function $U_{XC}$
[Fig.~\ref{figure3}(b)] and the uncorrected intersubband polarizability
$\chi_{12}^0 \equiv \Pi_{12}^0+\Pi_{21}^0$ [Fig.~\ref{figure3}(a)].
In the two-subband limit, the vertex-corrected irreducible response
function is given by $\chi_{12}^0 [1-U_{XC} \chi_{12}^0]^{-1}$, which
clearly has an instability when $\mid U_{XC} \chi_{12}^0 \mid \ge 1$
[cf. Eq.~(\ref{eq:intersubband})].
As shown in Fig.~\ref{figure3}(a), this ``Stoner criterion'' is satisfied
in the range of sheet densities in which the SDE has collapsed.

Having established the existence of the SDE-collapsed phase for a
particular DQW structure, we can vary the geometry of the structure
to study the persistence of this phase.
For well widths fixed at 139~\AA, the calculated zero-temperature phase
diagram in terms of the sheet density and barrier width $d_B$ is shown
in Fig.~\ref{figure4}.
For very small barrier widths, $\Delta_{SAS}$ is too large for the
vertex correction to overcome it, even at low densities,
making the normal phase the only stable phase.
For very large barrier widths, on the other hand, $\Delta_{SAS}$ is
exponentially small, and the normal phase gives way to the excitonic
one at extremely low densities.
Of course, for such exponentially small $\Delta_{SAS}$, the critical
temperature for our predicted instability is also exponentially low,
implying that the phase transition in large-$d_B$, low-density DQW
structures would be unobservable in practice.\cite{footnote}
At higher densities in the large-$d_B$ structures, the normal phase
re-asserts itself due to the diminishing influence of the vertex correction.
For intermediate values of the barrier width, we see the re-entrant
behavior described above with an upper critical density that is nearly
independent of the barrier width.
A similar phase diagram is obtained when the barrier width is fixed
at 40~\AA and the well widths are varied, as demonstrated by the
inset to Fig.~\ref{figure4}.
By expressing the sheet density and $\Delta_{SAS}$ in dimensionless
form, the data from Fig.~\ref{figure4} can be reduced to the universal
zero-temperature phase diagram shown in Fig.~\ref{figure5}.
From this figure, we see quite clearly that the excitonic phase
appears in a range of densities below a critical symmetric-antisymmetric
splitting $\Delta_{SAS}$ and that this density range increases as
$\Delta_{SAS}$ decreases.

The results discussed so far for the intersubband SDE energy are for
zero in-plane momentum transfer.
Fig.~\ref{figure6} shows the dispersion relation of the lowest
intersubband spin-density mode for several sheet densities $N_s$
above $N_C$.
The most important feature shown is that the SDE mode becomes soft at a 
finite value of the in-plane momentum transfer, $q_c$, at a critical density 
somewhat higher than the critical density found at zero in-plane momentum 
transfer.
The critical momentum is given by $q_c=k_{F1}-k_{F2}$, where $k_{Fi}$ is the 
in-plane Fermi wave vector of the $i$th subband in the normal state.
This indicates that the excitonic instability, for densities
slightly above $N_C$, may occur at a finite value of in-plane $q$,
a point to which we shall return in Secs.~\ref{sec:ns_instability} and
\ref{sec:conclusion}.

Thus, we expect that a softening of the intersubband SDE may be observed
either at fixed $q$ and varying $N_s$ [cf. Fig.~\ref{figure7}(a)]
or for $q \rightarrow q_c$ at certain fixed values of $N_s$
[cf. Fig.~\ref{figure7}(b)].
In Fig.~\ref{figure7}(a), we show the spectral function of the
intersubband spin-density excitations as the instability is approached
from the high-density side at very small $q$.
As $N_s$ approaches $N_C$, the spectral peak shifts towards zero energy 
and the line narrows.
On the other hand, the approach to the critical momentum transfer
$q_c$ at densities just above $N_C$, presented in Fig.\ \ref{figure7}(b) for
$N_s=N_C=0.7 \times 10^{11}$ cm$^{-2}$ at various values of $q$
approaching $q_c$, the line broadens as it softens.
These spectra are similar to what would be observed in Raman scattering
experiments in the cross-polarization
geometry.\cite{pin-abs,pinczuk,shanabrook,decca} 
For sheet densities within the SDE-collapsed phase, the TDLDA treatment
employed in the Section breaks down, and we must employ the techniques
of Sec.~\ref{sec:SCHF} to determine the collective modes, which we shall
do in Sec.~\ref{sec:AF}.

Finally, we mention that the instability discussed here for DQWs can also
happen in wide {\em single} quantum wells, in which the effective potential
develops a barrier in the center at moderate densities, becoming
similar to a DQW.\cite{he-das-xie}
An example of this phenomenon is illustrated in Fig.~\ref{figure8};
notice that the electron density profile is similar to that for the DQW
system shown in Fig.~\ref{figure1}.
In this {\em effective} double well system, the separation between the two 
lowest lying subbands is small, about 0.2~meV, as in the coupled DQWs, and
the possibility of a suppression of the intersubband SDEs by the
vertex correction arises.
A calculation of the SDE energies shows that this is indeed the case,
as can be seen in Fig.~\ref{figure9}.
For the square well considered, whose width is 1000~\AA, the region in
which the SDE has collapsed is somewhat smaller than that for the DQW of
Fig.~\ref{figure1} with a 40~\AA barrier width, but it occurs around the
same $N_s \sim 0.4 \times 10^{11}$ cm$^{-2}$.
Hence, inelastic light scattering experiments should be able to
detect the SDE instability in wide square wells also.

\section{Ferromagnetic Phase}
\label{sec:FM}

In the preceding Section, we showed that the
electron gas confined in a coupled double-quantum-well structure
has an instability which is indicated by the vanishing of the energy
of the intersubband spin-density excitations.
One of the goals of this Section is to establish whether
this instability corresponds to a phase transition from the normal 
spin-unpolarized (paramagnetic) ground state to a spin-polarized 
(ferromagnetic) one.
A second goal is to employ the LSDA formalism of Sec.~\ref{sec:LSDA} to
the ferromagnetic phase transition in single square quantum wells.
In particular, we are interested in the dependence of the critical
density on the well width, as it would reveal an aspect of the crossover 
from two- to three-dimensional behavior of the electron gas which has
not been explored in the past.
Our main finding is that the LSDA does predict a ferromagnetic transition
in single quantum wells and, moreover, that as the electron gas is widened 
the transition density decreases (the ferromagnetic phase becomes less 
favorable).
This dependence of the critical density on the electron-layer width 
agrees with the well-known fact that the influence of the Coulomb 
interaction is stronger for lower dimensionality.

We first determine whether the excitonic 
instability studied in Sec.~\ref{sec:instability} can be explained in 
terms of a ferromagnetic transition, that is, a transition from the usual 
spin-unpolarized ground state to a partially or fully polarized one.
We concentrate on a coupled double-quantum-well system in which each 
individual square well is 139~\AA wide and the well separation is 40~\AA
-- the main structure studied in Sec.~\ref{sec:instability}.
As in the better-known 3D case, a fully polarized phase
is expected at low density, and a normal, unpolarized phase
at high density.
To compute the spin polarization of the system at a given
sheet density $N_s$ within the iterative, self-consistent LSDA method, 
it should be enough, in principle, to introduce a slight asymmetry 
in the initial choice of spin densities.
If the correct ground state were unpolarized, the initial asymmetry would
rapidly disappear in the iteration process.
On the other hand, if a polarized state were expected, the initial
small polarization would increase until convergence to the fully
polarized is achieved.
However, since the solution of the self-consistent set of equations
of the LSDA method is affected by numerical inaccuracies, 
in practice our algorithm is sensitive to the initial guess for the
spin density profiles.
Therefore, we employ the following method to determine the 
ground-state polarization of the system.
For a given density $N_s$, we solve the self-consistent algorithm 
starting with the spin densities $n_1^+(z) = \eta \, n_u(z)$, 
$n_1^-(z) = (1-\eta) \, n_u(z)$, $n_2^+(z) = 0$, and $n_2^-(z) = 0$
for $\eta=0.55$ and $0.95$ and $n_u(z)$ the density from the unpolarized
LDA calculation (only one subband is occupied at the low $N_s$ studied).
If the calculation converges to a polarized or an unpolarized state for 
both values of $\eta$, we take the result as the true polarization at that
given $N_s$. 
Otherwise, we assume that the result is affected by the insufficient
numerical precision.

With this method, we obtain the phase diagram of spin polarization
as a function of $N_s$ for the coupled DQW defined above.
The calculation converges to a fully polarized state, regardless
of the initial choice of spin densities, for
$N_s < 4 \times 10^{8}$ cm$^{-2}$ and to an unpolarized state for
$N_s > 1.7 \times 10^{9}$ cm$^{-2}$.
In the range $N_s=$ 0.4--1.7 $\times 10^{9}$ cm$^{-2}$, the solution
is polarized for the choice $\eta=0.95$, and unpolarized for 
$\eta=0.55$.

Therefore, our first conclusion is that, within the LSDA,
{\it there is a ferromagnetic transition} as a function of
the electron density $N_s$ in a coupled DQW system;
for our particular choice of parameters, the transition occurs around
$N_s \approx 1 \times 10^{9}$ cm$^{-2}$ at zero temperature.
On the other hand, the ferromagnetic transition occurs
at a density which is almost two orders of magnitude smaller 
than the critical density $N_C$ of the excitonic transition of
Sec.~\ref{sec:instability}.
Moreover, it does not seem to show re-entrant behavior at
lower density as the excitonic transition does.
Based on these differences, we conclude that {\it the ferromagnetic
transition cannot be identified as the excitonic phase transition}
of Sec.~\ref{sec:instability}.

Having achieved the first aim of this Section, we now turn to the
second, namely the possibility of ferromagnetism in single square wells.
We wish to determine the critical density of the ferromagnetic transition
as a function of the well width in order to take the first step in
studying the evolution of this transition as a function of the
dimensionality of the electron gas.
The analytical results for two and three dimensions summarized in 
Sec.~\ref{sec:LSDA} indicate that, in 2D, the exchange energy
is more important than in 3D, indicating a higher 2D critical density.
In quasi-2D systems, we therefore expect that the ferromagnetic critical
density should decrease as the well width is increased.

Employing the method described above, we study the ferromagnetic
transition in five single square wells of widths 
$d_W = a_0, \, 2a_0, \, 4a_0, \, 6a_0$, and $8 a_0$,
where $a_0$ = 98.7~\AA is the effective Bohr radius in GaAs.
The resulting phase diagram is presented in Fig.~\ref{figure10}
in terms of $N_s$ versus well width, and also in 
terms of the 2D and 3D $r_s$ and well width;
the vertical bars give the density range where the polarization of
our solution depends on the initial choice of spin densities.
As expected, we see that the critical density decreases with
increasing well width. 
However, the limiting values of $r_s$ for narrow and wide wells
cannot be directly compared with the pure two- and three-dimensional 
Hartree-Fock values for three reasons.
First, the analytical results are obtained in the jellium model, whereas 
in our quantum well calculations the positive charges of the ionized donors 
are located far away from the electron gas, which should produce an 
important change in the direct Coulomb energy and hence affect the 
ferromagnetic critical density.
Second, our calculation includes correlation effects which go beyond the 
exchange interaction. 
And the third reason to expect differences is that, in the LSDA, exchange 
is treated in a local and static approximation.

This exchange-correlation potential is obtained from the ground-state
energy of a uniform electron gas, which is not known exactly and must
be calculated within some approximation scheme.
Several authors have proposed different parametrizations of
the exchange-correlation potential, which are based on different
calculations of the 3D-electron-gas ground-state energy and which
therefore differ somewhat from each other.
We have checked some of our results using the parametrization of
Gunnarsson and Lundqvist,\cite{gun-lun} which gives a larger difference 
between the potentials for different spin polarizations
than the Ceperley-Alder one, making the ferromagnetic phase
more favorable.
For example, for the square well of width $d_W =4a_0$, the 
critical density is $1.3 \times 10^{10}$ cm$^{-2}$, ten times 
higher than with Ceperley-Alder.
Thus, our calculations indicate that within the LSDA a ferromagnetic
transition is indeed present in the quasi-two-dimensional electron gas,
but the value of the critical density for the transition cannot be
accurately determined with this method.

\section{Antiferromagnetic Phase}
\label{sec:AF}

Based on the results of the preceding two subsections, we can
draw the following conclusions:  (1) double or wide single quantum
well structures can undergo an excitonic instability in a range
of densities, and (2) this instability cannot be associated with
a ferromagnetic transition.
We would like to understand the nature of the ground state of the
excitonic phase and determine its excitations, but we see that the
density functional approach cannot assist us further.
In order to make progress, we employ the self-consistent Hartree-Fock
formalism developed in Sec.~\ref{sec:SCHF}, which allows us
to study ground states with broken symmetries which are not accessible
through ordinary density functional theory.

As pointed out in Sec.~\ref{sec:SCHF}, the self-consistent
Hartree-Fock model for systems with a Coulomb interaction is
not quantitatively accurate but should reproduce the qualitative
features of the new phase.
Our goal is therefore to construct a minimal model of the systems
which exhibit the excitonic instability which is still tractable.
We can make several observations about the excitonic instability which
will guide us in developing this model.
First, the instability in the LDA calculations occurs
with a {\it wave-vector independent} interaction,
as seen from the LDA expressions for the irreducible
polarizability, Eqs.~(\ref{eq:pi-irred}) and (\ref{eq:UXC}).
Second, the instability is signaled by a soft mode which is
an excitation between the lowest two subbands and is accompanied
by a spin flip.
Thus, our minimal model will focus only on the lowest
two subbands, allow for off-diagonal terms in the subband
and spin indices, and use an interaction which in constant in
wave-vector space.
In the rest of this Section, we discuss the results which follow
from this minimal model; a preliminary report of this work has
already appeared in the literature.\cite{rad-das}

\subsection{Point-Contact Model}
\label{sec:model}

A {\bf q}-indepent interaction has several simplifying consequences.
First, as seen from Eq.~(\ref{eq:Sigma}), the self-energy will also
be wave-vector independent.
From Eq.~(\ref{eq:eigen_eqn}), this result implies that the
eigenfunctions $\varphi^c_a({\bf k})$ are also wavevector independent
and that the interacting eigenenergies may be written
\begin{equation}
E^c ({\bf k}) =
  e^c + \frac{\hbar^2 k^2}{2 m^*} - \mu.
\label{eq:Ec}
\end{equation}
Because these eigenenergies are the energies of the interacting
quasiparticles, the effect of the interaction is to shift and/or
rearrange the non-interacting bands without destroying their
parabolic dispersion.
Consequently, the broken-symmetry phases available within this model
will not exhibit any energy gap, in contrast to what occurs in
superconductivity\cite{Schrieffer} and in other excitonic transitions
which have been studied.\cite{Rice}
The reason for this difference is that the interacting quasiparticles are
linear combinations of electrons and holes in the latter cases, whereas
in this case they are linear combinations of electrons from different
subbands in the former.

Second, the simplest interaction which is {\bf q}-independent
is a delta-function in real space:
$V({\bf R}) = V \delta ({\bf R})$.
Inserting this form into Eq.~(\ref{eq:V}) yields
\begin{equation}
V_{ab,cd} ({\bf q}) =
  \delta_{\sigma_a \sigma_b} \delta_{\sigma_c \sigma_d} \,
  V_{n_a n_b, n_c n_d}
\end{equation}
with
\begin{equation}
V_{n_a n_b, n_c n_d} =
  V \, \int dz \, \xi_{n_a}^{*} (z) \, \xi_{n_b}^{~} (z) 
  \, \xi_{n_c}^{*} (z) \, \xi_{n_d}^{~} (z) .
\label{eq:Vn}
\end{equation}
Making the standard assumption that the quantum well structure is
symmetric, the wave functions $\xi_n (z)$ may be
chosen to be real and parity eigenstates.
From Eq.~(\ref{eq:Vn}), these symmetries imply that the order of
the subband indices in the matrix elements is irrelevant and that
matrix elements of the form $V_{11,12}$, $V_{12,22}$, and cyclic
permutations vanish.
Thus, there are only three independent matrix elements:
$V_{11} = V_{11,11}$, $V_{22} = V_{22,22}$, and
$V_{12} = V_{11,22}$ + permutations.
Since we are interested in the minimal model exhibiting the
excitonic instability, we will neglect $V_{11}$ and $V_{22}$
in order to focus on the intersubband effects, leaving a model
with two free parameters:  the sheet density $N_s$ and the
intersubband interaction $V_{12}$.

Applying these approximations to Eq.~(\ref{eq:Sigma2}), we find that the
self-energy reduces to
\begin{equation}
\Sigma_{n_1 \sigma_1, n_2 \sigma_2} =
  V_{n_1 n_2, n_3 n_4} n^c \left\{ \begin{array}{ll}
  \varphi_{n_3 -\sigma_1}^{c} \varphi_{n_4 -\sigma_1}^{*c}, &
  \sigma_1 = \sigma_2 \\
  - \varphi_{n_3 \sigma_1}^{c} \varphi_{n_4 \sigma_2}^{*c}, &
  \sigma_1 \neq \sigma_2 \end{array} \right. .
\label{eq:Sigma_pc}
\end{equation}
This equation is supplemented by the band-filling constraint,
Eq.~(\ref{eq:n}), with the electron density in each interacting
subband given by
\begin{equation}
n^c = \frac{1}{A} \sum_{\bf k} f( E^c ({\bf k}) ).
\end{equation}

In addition, the integral equations determining the polarizability
[Eqs.~(\ref{eq:Pi})-(\ref{eq:gamma})] can be reduced to a matrix equation.
The first step in this reduction is to observe that the vertex
function $\gamma$ in Eq.~(\ref{eq:gamma}) is a function only
of $q_n$ when the the interaction has no ${\bf q}$-dependence
and satisfies the equation
\begin{equation}
\gamma_{ab,cd} (q_n) = 
  \delta_{ac} \delta_{bd} +
  \left[ V_{ba,ef} - V_{bf,ea} \right] \, \Pi_{ef,gh}^{(0)} (q_n) \,
  \gamma_{gh,cd} (q_n)
\label{eq:gamma-noq}
\end{equation}
where
\begin{equation}
\Pi_{ab,cd}^{(0)} (q_n) =
  \frac{T}{A} \sum_{k_m} G_{ca} (k_m) \, G_{bd} (k_m + q_n)
\label{eq:Pi0}
\end{equation}
is the polarizability without vertex corrections.
In the interacting system, $\Pi_{ab,cd}^{(0)}$ may be written through the
use of Eq.~(\ref{eq:Gspec}) as
\begin{equation}
\Pi_{ab,cd}^{(0)} (q_n) =
  \varphi_a^{*e} \varphi_b^{f} \varphi_{c}^{e} \varphi_{d}^{*f} \,
  \Pi^{ef} (q_n).
\label{eq:Pi0_phi}
\end{equation}
After analytically continuing to real frequencies, the function
$\Pi^{ef} (q)$ is just Eq.~(\ref{eq:pi0}) with $g_s = 1$ and using
the interacting energies [Eq.~(\ref{eq:Ec})].
Inserting this result into Eq.~(\ref{eq:Pi}) and rearranging yields
\begin{equation}
\left[ \delta_{ag} \delta_{bh} - \Pi_{ab,ef}^{(0)} (q) \,
  \left( V_{fe,gh} - V_{fh,ge} \right) \right] \, \Pi_{gh,cd} (q) =
  \Pi_{ab,cd}^{(0)} (q).
\label{eq:Pi_cm}
\end{equation}
Inverting this matrix equation gives the polarizability.
When this equation cannot be inverted for a particular $({\bf q},\omega)$,
i.e. when the determinant of the matrix in brackets vanishes, the
system exhibits a collective mode.
The polarization of these collective modes can be determined by,
for example, the method discussed in Sec.~\ref{sec:SCHF}.

Before proceeding to the discussion of the normal-state properties
of our point-contact model, it is appropriate to discuss briefly
the question of how well this model approximates reality.
The Hartree-Fock approximation is known to be a poor one in the
interacting electron gas because it neglects screening effects, but
including these effects realistically is a difficult problem which
has not yet been resolved.
In computing the collective modes, this problem is amplified by
the distinction one should draw between the interaction between
the bubbles, which is unscreened, and the interaction within the
bubbles that gives the non-trivial part of the vertex equation,
which should be screened.
Summing a particular set of screening diagrams may reduce the
error introduced into the self-energy, but would render the
collective mode calculation completely intractable.
Thus, we adopt a strong approximation to the actual Coulomb
interaction, a point-contact interaction, which should reproduce
the qualitative features of the screening effects while
leaving a solvable set of equations.

\subsection{Normal-State Instability}
\label{sec:ns_instability}

Our goal is to use the point-contact interaction model we have just
described as the minimal model for the excitonic instability.
It is therefore critical to verify that this model exhibits the instability
in its normal state, an exercise which will also provide guidance to
the nature of the broken-symmetry state.
The first step in this process is to compute the self-energy and
chemical potential in the interacting system.
In the normal paramagnetic phase of this model, we know that
the wave functions of the subbands are not mixed
and that the band structure consists of two parabolic subbands,
each with degenerate spin-up and spin-down components.
Consequently, the self-energy is diagonal in subband and spin
indices, and the eigenvectors of Eq.~(\ref{eq:eigen_eqn}) are
given by $\varphi_a^{c} = \delta_a^{c}$ with the corresponding
eigenvalues satisfying
$e^{1\uparrow} = e^{1\downarrow} \equiv e_1$ and
$e^{2\uparrow} = e^{2\downarrow} \equiv e_2$.
It then follows that $n^{1\uparrow} = n^{1\downarrow} \equiv n_1$,
$n^{2\uparrow} = n^{2\downarrow} \equiv n_2$,
$\Sigma_{1\uparrow,1\uparrow} = \Sigma_{1\downarrow,1\downarrow}
\equiv \Sigma_1$, and
$\Sigma_{2\uparrow,2\uparrow} = \Sigma_{2\downarrow,2\downarrow}
\equiv \Sigma_2$.
Inserting these formulae into Eq.~(\ref{eq:Sigma_pc}) yields the
self-energy equations $\Sigma_1 = V_{12} n_2$ and
$\Sigma_2 = V_{12} n_1$.

At finite temperatures, the simultaneous solution of these self-energy
equations and the band-filling constraint Eq.~(\ref{eq:n})
must be performed numerically, but at $T = 0$ the solution can be
obtained analytically.
In order to exhibit the zero-temperature solutions in a compact form,
we define the non-interacting subband splitting in terms of the
eigenvalues of Eq.~(\ref{eq:xidef}):
$\Delta_{\rm SAS}^{0} \equiv \epsilon_2 - \epsilon_1$.
The interaction renormalizes this splitting without
modifying the identity of the underlying subbands.
Applying the self-energy equations allows us to write this
renormalized splitting as
\begin{eqnarray}
\Delta_{\rm SAS}^{*} &\equiv& e_2 - e_1 \nonumber \\
  &=& \Delta_{\rm SAS}^{0} + V_{12} (n_1 - n_2).
\label{eq:Delta_star}
\end{eqnarray}
If the chemical potential is measured from the bottom of the lower
subband and $\mu < \Delta_{\rm SAS}^*$, then only the lower subband is
occupied at zero temperature, and we shall refer to this state as the
$N_1$ phase.
Alternatively, if $\mu > \Delta_{\rm SAS}^*$, both subbands are occupied,
and we shall refer to this as the $N_2$ phase.
By solving the equations for $\Delta_{\rm SAS}^*$ and the band filling
constraint Eq.~(\ref{eq:n}) simultaneously, we obtain
\begin{equation}
\Delta_{\rm SAS}^* = \left\{ \begin{array}{ll}
  \Delta_{\rm SAS}^0 + \frac{1}{2} N_s V_{12} & (N_1 \ \mbox{phase}) \\
  \frac{\Delta_{\rm SAS}^0}{1 - N_0 V_{12}} & (N_2 \ \mbox{phase})
  \end{array} \right.
\end{equation}
and
\begin{equation}
\mu = \left\{ \begin{array}{ll}
  \frac{N_s}{2 N_0} & (N_1 \ \mbox{phase}) \\
  \frac{N_s}{4 N_0} + \frac{1}{2} \frac{\Delta_{\rm SAS}^0}{1 - N_0 V_{12}}
    & (N_2 \ \mbox{phase})
  \end{array} \right.
\end{equation}
The crossover from $N_1$ to $N_2$ occurs when $\mu = \Delta_{\rm SAS}^*$
or $N_s / 2 N_0 \Delta_{\rm SAS}^0$ = $1 / (1 - N_0 V_{12})$.

With the self-energy and chemical potential computed, we are now in
a position to study the collective intersubband spin-density excitations
of this model to see if the excitonic instability occurs.
The condition for these excitations obtained from
Eq.~(\ref{eq:Pi_cm}) through the use of Eq.~(\ref{eq:Pi0_phi}) is
\begin{equation}
1 \pm {\rm Re} \left[ \Pi^{12} (q,\omega) + \Pi^{21} (q,\omega) \right]
  V_{12} = 0,
\label{eq:cm_pc}
\end{equation}
where the upper (lower) sign is for the spin (charge) density excitation
and we have dropped the spin indices since $\Pi^{ab}$ does not depend
on them in the normal state.
At ${\bf q} = {\bf 0}$ and $T = 0$, an analytic solution for the
frequency of the collective modes $\omega_0$ is possible which is
just the Ando result\cite{ando76} with a renormalized subband splitting:
\begin{equation}
\omega_0^2 = \left( \Delta_{\rm SAS}^* \right)^2 \mp
  2 V_{12} (n_1 - n_2) \Delta_{\rm SAS}^*.
\end{equation}
It is clear from this expression that the spin-density excitation (SDE)
will soften completely when $\Delta_{\rm SAS}^* \le 2 V_{12} (n_1 - n_2)$.
At zero temperature, the relations derived in the preceding paragraphs
can be used to show that the softening
occurs in the $N_1$ phase when $N_s V_{12} / 2 \Delta_{\rm SAS}^0 \ge 1$,
and in the $N_2$ phase when $N_0 V_{12} \ge \frac{1}{2}$.
These boundaries are shown in Fig.~\ref{fig:phase} along with the
line separating the $N_1$ and $N_2$ phases.
The shaded area represents the region in which the spin-density
excitation is soft at $q = 0$ in this model.
The phase corresponding to the soft mode occupies a
large region of parameter space and obtains at any density providing
the interaction $V_{12}$ is sufficiently strong.

At non-zero wave vector, the Eq.~(\ref{eq:cm_pc}) must be solved
numerically.
From the low-density, $N_1$ phase, the results of this calculation
are shown in Fig.~\ref{fig:A1-A2} for increasing density at fixed
interaction $V_{12}$.
As illustrated by the figure, the $q = 0$ spin-density excitation
softens with increasing density and vanishes beyond a critical
density which depends on the interaction strength $V_{12}$.
At higher densities, the small-$q$ solutions to Eq.~(\ref{eq:cm_pc})
become imaginary, and the largest imaginary frequency -- indicating
the most unstable wave vector -- continues to be at $q = 0$.
Thus, the transition from the low-density side is very clearly a
$q = 0$ instability, as is also seen in the TDLDA calculations.

The situation from the high-density, $N_2$ side is more complicated.
The dispersion curves for this case computed from Eq.~(\ref{eq:cm_pc})
at fixed density and increasing interaction strength closely resemble
those in Fig.~\ref{figure6}.
As the interaction strength $V_{12}$ increases, the energy of the SDEs
are reduced until the entire dispersion curve collapses along
a range of wave vectors from $q = 0$ to $k_{F1} - k_{F2}$, the difference
of the Fermi wave vectors of the two subbands, at a critical $V_{12}$.
The reason for this collapse is seen from Eq.~(\ref{eq:cm_pc}) when
one observes that at $T = 0$ in the $N_2$ phase
\begin{equation}
{\rm Re} \left[ \Pi^{12} (q,\omega) + \Pi^{21} (q,\omega) \right] = - 2N_0
\label{eq:Pi_N2}
\end{equation}
for $q < k_{F1} + k_{F2}$, indicating that if the energy of a SDE
with one of these wave vectors vanishes, they all must vanish.
At interaction strengths slightly larger than that required for a collapse,
the solutions to Eq.~(\ref{eq:cm_pc}) are imaginary in this range
of wave vector with the largest imaginary frequency occurring at $q = 0$.
These calculations suggest a $q = 0$ transition from the high-density
side, but the density-functional computations in
Sec.~\ref{sec:instability} find a softening at the non-zero
wave vector $q = k_{F1} - k_{F2}$.
The discrepancy arises from the fact that the calculations in
Sec.~\ref{sec:instability} include all the subband levels, whereas
the model in this Section contains only two.
The inclusion of higher subbands, even if they are unoccupied, disrupt
the cancellations leading to Eq.~(\ref{eq:Pi_N2}) and yield a
$q$-dependent function which will select some wave vector.
The consequences of a finite ordering vector will be discussed in
Sec.~\ref{sec:conclusion} but are beyond the scope of our simple model.
We nonetheless expect that many of the qualitative insights from our
model will apply to more comprehensive theories.

One of the most important insights that can be gained by the study of
the normal-state instability is a physical intuition about
the nature of the ground state in the region of the the phase
diagram in which the SDE has softened.
In structural phase transitions, a particular phonon softens, and
the polarization of that phonon determines the new structure.
In the same way, the polarization of the soft spin-density excitation
should reveal a great deal about the new ground state.
This polarization can be computed by means of
Eqs.~(\ref{eq:polarization})-(\ref{eq:fext}), which leads to a density
response in the normal state given by
\begin{equation}
\left< \rho^{\mu} ({\bf R}, t) \right> \propto
  e^{i({\bf q \cdot r} - \omega t)} \, \xi_1 (z) \, \xi_2 (z) \,
  \left[ \Pi^{12} (q,\omega) + \Pi^{21} (q,\omega) \right]
  (\hat{x} - i\hat{y}) ,
\end{equation}
which is shown graphically in Fig.~\ref{fig:polarization}(a).
Notice that the response occurs in the spin channel
and has the character of a traveling wave transverse to the
layering direction.
Also observe that, since the wave function of the lowest subband
$\xi_1 (z)$ is even in $z$ due to the assumed symmetry of the
quantum well structure, and the wave function of the next subband
$\xi_2 (z)$ is odd, the overall response odd in $z$.
Thus, the intersubband spin density excitation corresponds to a
spin-density wave in which the spin density is antiferromagnetically
correlated between the quantum wells.

With this interpretation, the nature of the excitonic instability
becomes obvious:
the softening of the $q = 0$ SDE corresponds to the formation
of antiferromagnetic order between the quantum wells with no modulation
of the spin density within a well.
A profile of the resulting spin density is presented in the
inset of Fig.~\ref{fig:AFop}.
Note that there are three degenerate SDEs in the normal phase which
may be associated with spin polarizations along the three Cartesian
directions.
Although all three soften simultaneously, we expect that a particular
spin direction will be selected by the system as in the ferromagnetic
transition, but there is {\it a priori} no restriction of the direction
of the spin polarization selected.

To summarize the results of the application of our simple model
to the normal state, we find a softening of the intersubband
spin-density excitations at $q = 0$ for a wide range of model parameters.
These collective excitations produce a spin-density modulation which is
antiferromagnetically correlated between the wells.
Thus, the softening of the $q = 0$ spin density modes indicate the
formation of an antiferromagnetic phase in which the spins in both
wells are polarized but in opposite directions.
In the more realistic density-functional calculations of
Sec.~\ref{sec:instability}, a $q \neq 0$ instability was
indicated which would imply an additional intra-well modulation of
the spin density (see Sec.~\ref{sec:conclusion}).
For simplicity, we focus on the $q = 0$ phase.

\subsection{Antiferromagnetic Ground State and Thermodynamics}
\label{sec:AFgs}

With an understanding of the nature of the excitonic instability as
an antiferromagnetic ordering, we are able to examine the new ground
state and its properties within our model.
We first note that both spin rotation invariance and parity are
broken in the antiferromagnetic state, so we are studying a genuine
symmetry-breaking phase transition.
The first task encountered in studying
such a transition is to identify the order parameter.
For concreteness, consider the antiferromagnetic phase with the
spin density aligned along the $\hat{x}$ direction.
From Eqs.~(\ref{eq:rho_r}) and (\ref{eq:rho_a}), the only non-zero
component of the spin density is then
\begin{equation}
\left< s^{x} (z) \right> =
  \frac{\hbar}{2} \, \xi_{n_1} (z) \, \xi_{n_2} (z) \,
  \left< \rho_{n_1 \uparrow, n_2 \downarrow} ({\bf q} = {\bf 0}) \right>.
\end{equation}
Because the antiferromagnetism requires that $\left< s^{x} (z) \right>$
be odd in $z$, only the terms which are off-diagonal in subband
index are non-zero in this phase.
Defining
\begin{equation}
\rho_{n_1 n_2}^{\mu} =
  \sigma_{\sigma_1 \sigma_2}^{\mu} \sum_{\bf k}
  \left< c_{n_1{\bf k}\sigma_1}^{\dag} c_{n_2{\bf k}\sigma_2}^{~} \right>
\end{equation}
and using Eq.~(\ref{eq:rho_c}), this implies that
\begin{eqnarray}
\left< s^{x} (z) \right> &=&
  \frac{\hbar}{2} \, \xi_{n_1} (z) \, \xi_{n_2} (z) \, \sum_{\bf k} \left[
  \left< c_{1{\bf k}\uparrow}^{\dag} c_{2{\bf k}\downarrow}^{~} \right> +
  \left< c_{2{\bf k}\uparrow}^{\dag} c_{1{\bf k}\downarrow}^{~} \right>
  \right] \\
&=&   \frac{\hbar}{2} \, \xi_{n_1} (z) \, \xi_{n_2} (z) \,
  \left[ \rho_{12}^{x} + \rho_{21}^{x} \right] .
\label{eq:OP}
\end{eqnarray}
$\rho_{12}^{x} + \rho_{21}^{x}$ is finite in the antiferromagnetic phase
and vanishes in the normal phase, so it is the order parameter of the
phase transition.
From the spin-rotational invariance of the system, the direction in
which the spin density can be polarized in the antiferromagnetic phase
is arbitrary, so the general form of the order parameter is
\begin{equation}
N^i = \rho_{12}^{i} + \rho_{21}^{i}, \, i = 1,2,3.
\label{eq:OP_N}
\end{equation}

With this identification, we can construct a form for the self-energy
matrix which permits the off-diagonal expectation values we require
yet preserves the other symmetries of the system.
This {\it ansatz} may be written\cite{num-note}
\begin{equation}
\Sigma_{ab} = \left[ \begin{array}{cccc}
  \Sigma_1 & 0 & 0 & \Sigma_{\rm od} \\
  0 & \Sigma_1 & \Sigma_{\rm od} & 0 \\
  0 & \Sigma_{\rm od} & \Sigma_2 & 0 \\
  \Sigma_{\rm od} & 0 & 0 & \Sigma_2
  \end{array} \right] .
\end{equation}
Inserting this form into the eigenvalue equation Eq.~(\ref{eq:eigen_eqn})
yields the eigenvectors
\begin{equation}
\varphi_a^c = \left[ \begin{array}{cccc}
  u & 0 &  0 & -v \\
  0 & u & -v &  0 \\
  0 & v &  u &  0 \\
  v & 0 &  0 &  u
  \end{array} \right]
\label{eq:evec}
\end{equation}
(columns correspond to the superscript) and eigenenergies
$e^c$ = $\{ e^{+}, e^{+}, e^{-}, e^{-} \}$ [cf. Eq.~(\ref{eq:Ec})], where
\begin{equation}
\left( \begin{array}{c} u \\ v \end{array} \right) =
  \left[ \frac{1}{2}
  \left( 1 \mp \frac{\Delta_{\rm SAS}^0 + \Sigma_2 - \Sigma_1}{2 D}
  \right) \right]^{1/2},
\label{eq:uv}
\end{equation}
\begin{equation}
e^{\pm} = \frac{\Delta_{\rm SAS}^0 + \Sigma_2 + \Sigma_1}{2} \pm D,
\label{eq:eval}
\end{equation}
and
\begin{equation}
D = \left[
  \left( \frac{\Delta_{\rm SAS}^0 + \Sigma_2 - \Sigma_1}{2} \right)^2
  + \Sigma_{\rm od}^2 \right]^{1/2}.
\label{eq:D}
\end{equation}
The eigenenergy spectrum indicates that the single-particle band structure
in the antiferromagnetic phase consists of two sets of two degenerate
parabolic subbands separated in energy by
$\Delta_{+-} = e^{+} - e^{-} = 2D$.
This band structure is similar to that in the normal phases, but the
wave functions of the interacting quasiparticles are considerably different:
from Eq.~(\ref{eq:evec}), we see that these quasiparticles are linear
combinations of electrons from the two non-interacting subbands.

The parameters in the eigenenergies and eigenvalues are determined
from the self-consistency equations obtained by substituting the
eigenvectors into reduced self-energy equation, Eq.~(\ref{eq:Sigma_pc}).
This procedure gives
\begin{eqnarray}
\Sigma_1 &=& V_{12} \left( n^{-} u^2 + n^{+} v^2 \right),
  \label{eq:S1} \\
\Sigma_2 &=& V_{12} \left( n^{+} u^2 + n^{-} v^2 \right),
  \, \mbox{and} \label{eq:S2} \\
\Sigma_{\rm od} &=& 2 V_{12} \left( n^{-} - n^{+} \right) uv .
  \label{eq:Sod}
\end{eqnarray}
Using the definitions of $u$ and and $v$ [Eq.~(\ref{eq:uv})] and the
band filling constraint
\begin{equation}
N_s = 2 \left( n^{-} + n^{+} \right),
\label{eq:Ns}
\end{equation}
these self-energy equations can be written
\begin{eqnarray}
\Sigma_1 + \Sigma_2 &=& \frac{1}{2} V_{12} N_s,
  \label{eq:Splus} \\
\Delta_{\rm SAS}^0 + \Sigma_2 - \Sigma_1 &=&
  \frac{\Delta_{\rm SAS}^0}{1 - V_{12} ( n^{-} - n^{+} ) / 2D},
  \, \mbox{and} \label{eq:Sminus} \\
\Sigma_{\rm od} &=& \frac{V_{12} ( n^{-} - n^{+} )}{D} \, \Sigma_{\rm od}.
  \label{eq:Sod2}
\end{eqnarray}
In the antiferromagnetic phase, $\Sigma_{\rm od} \neq 0$, implying
from Eq.~(\ref{eq:Sod2}) that $D = V_{12} ( n^{-} - n^{+} )$.
Eq.~(\ref{eq:Sminus}) then becomes
$\Delta_{\rm SAS}^0 + \Sigma_2 - \Sigma_1 = 2 \Delta_{\rm SAS}^0$,
from which we obtain $D^2 = (\Delta_{\rm SAS}^0)^2 + \Sigma_{\rm od}^2$
by Eq.~(\ref{eq:D}).
Comparing the two expressions for $D$ yields
\begin{equation}
\Sigma_{\rm od} =
  \left[ V_{12}^2 ( n^{-} - n^{+} )^2 - (\Delta_{\rm SAS}^0)^2 \right]^{1/2}.
\label{eq:Sigma_od}
\end{equation}
The splitting between the interacting bands is
\begin{equation}
\Delta_{+-} = 2D = 2 V_{12} ( n^{-} - n^{+} )
\label{eq:Delta_pm}
\end{equation}
and the chemical potential is determined implicitly by Eq.~(\ref{eq:Ns}).

At zero temperature, these equations can be easily solved.
First, suppose $n^{+} = 0$.
Then Eqs.~(\ref{eq:Ns}) and (\ref{eq:Sigma_od} imply that
\begin{equation}
\Sigma_{\rm od} =
  \left[ \left( \frac{V_{12} N_s}{2} \right)^2 -
  (\Delta_{\rm SAS}^0)^2 \right]^{1/2},
\label{eq:Sigma_od0}
\end{equation}
Eq.~(\ref{eq:Delta_pm}) becomes $\Delta_{+-} = V_{12} N_s$, and
the band-filling constraint yields $N_s = 2 N_0 \mu$.
This solution is consistent if
(1) $V_{12} N_s / 2 \Delta_{\rm SAS}^0 \ge 1$ [Eq.~(\ref{eq:Sigma_od0})], and
(2) $\mu \le \Delta_{+-}$, which is equivalent to $N_0 V_{12} \ge 1/2$.
These constraints are precisely those obtained in
Sec.~\ref{sec:ns_instability} as the boundaries of the normal-state
instability and are depicted in Fig.~\ref{fig:phase},
demonstrating that $n^{+} = 0$ over the entire range of the
antiferromagnetic phase.
Thus, only the lowest, ``-''  bands are occupied in this phase.
From Eq.~(\ref{eq:Sigma_od0}), we also observe that
$\Sigma_{\rm od}$ rises from zero continuously along the boundary
shared with the $N_1$ phase [cf. Fig.~\ref{fig:phase}],
indicating a second-order phase transition at zero temperature.
Along the boundary with the $N_2$ phase, however, $\Sigma_{\rm od}$
jumps discontinuously to a finite value, showing that this transition
is {\it first}-order at $T = 0$.

This interpretation is confirmed by a calculation of the energy change
of the system across the transition, which also demonstrates the stability
of the antiferromagnetic phase relative to the normal phases.
From the relations
\begin{eqnarray}
\frac{1}{A} \sum_{\bf k} f \left( \frac{\hbar^2 k^2}{2 m^*} - x \right)
 &=& N_0 T \ln \left( e^{x / T} + 1 \right), \label{eq:n_finiteT} \\
 &\stackrel{\longrightarrow}{\mbox{\small T $\rightarrow$ 0}}&
 N_0 x \, \theta (x) \label{eq:n_T0}
\end{eqnarray}
and
\begin{equation}
\frac{1}{A} \sum_{\bf k} \frac{\hbar^2 k^2}{2 m^*} \,
  f \left( \frac{\hbar^2 k^2}{2 m^*} - x \right)
 \stackrel{\longrightarrow}{\mbox{\small T $\rightarrow$ 0}}
 N_0 \frac{x^2}{2} \, \theta (x) ,
\label{eq:T_T0}
\end{equation}
and Eqs.~(\ref{eq:Ec}), (\ref{eq:ea}), (\ref{eq:ON}), and (\ref{eq:n}),
we can write the total energy [Eq.~(\ref{eq:Ephi})] as
\begin{equation}
\frac{E}{A} = \frac{n^c}{2} \left(
  \left| \varphi^{c}_a \right|^2 \epsilon_{n_a} + \mu \right) .
\label{eq:Euse}
\end{equation}
Inserting the subband occupations, eigenvectors, and chemical
potential obtained from the solution of the self-energy and
band-filling equations at $T = 0$, we find for $N_1$ phase
\begin{equation}
\frac{E^{N_1}}{A} = \frac{N_s^2}{4 N_0},
\label{eq:E_N1}
\end{equation}
for the $N_2$ phase
\begin{equation}
\frac{E^{N_2}}{A} =
  \frac{N_s^2}{8 N_0} \left( 1 + N_0 V_{12} \right) +
  \frac{\Delta_{\rm SAS}^0}{2} \left( N_s +
  \frac{N_0 \Delta_{\rm SAS}^0}{1 - N_) V_{12}} \right),
\label{eq:E_N2}
\end{equation}
and for the antiferromagnetic (AF) phase
\begin{equation}
\frac{E^{AF}}{A} =
  \frac{N_s^2}{4 N_0} - \frac{1}{2 N_0 V_{12}} \,
  \left( \frac{N_s V_{12}}{2 \Delta_{\rm SAS}^0} - 1 \right)^2 .
\label{eq:E_AF}
\end{equation}
We see immediately from Eqs.~(\ref{eq:E_N1}) and (\ref{eq:E_AF})
that, in the range of parameter space in which
both $N_1$ and $AF$ solutions exist, $E^{AF} < E^{N_1}$.
In addition, the energy at the $N_1$-$AF$ phase boundary
[cf. Fig.~\ref{fig:phase}] can be seen to be continuous and with
continuous first derivatives.
The second derivative is discontinuous at the phase boundary, however,
showing that the phase transition is second order.
Similarly, it can be shown that Eqs.~(\ref{eq:E_N2}) and (\ref{eq:E_AF})
yield $E^{AF} < E^{N_2}$ in the region of parameter space where both
solutions exist and that the {\it first} derivative of the energy difference at
the phase boundary is discontinuous.
Thus, the antiferromagnetic phase is energetically stable with
respect to the normal phases whenever the broken-symmetry solution exists,
and the phase transition is second order from the $N_1$ phase and first
order from the $N_2$ phase at zero temperature in mean-field theory.

At finite temperature, the self-energy equations in combination with
the band-filling constraint must be solved numerically.
The results of these computations can be used to obtain several
quantities which characterize the antiferromagnetic phase:
the transition temperature $T_c$ and the temperature dependence of
the order parameter $N^x = \left( \rho_{12}^x + \rho_{21}^x \right)$
and the specific heat $c_V$.
The temperature at which the antiferromagnetic transition occurs
is found by linearizing the self-energy
equations in the off-diagonal self-energy $\Sigma_{\rm od}$.
This procedure allows Eq.~(\ref{eq:Sod2}) to be written as
\begin{equation}
V_{12} (n^{-} - n^{+}) = \Delta_{\rm SAS}^0
\label{eq:Tc_od}
\end{equation}
and also leads to the result $\Delta_{+-} = \Delta_{\rm SAS}^*$ for
$\Delta_{\rm SAS}^*$ defined by Eq.~(\ref{eq:Delta_star}).
From these two expressions, we deduce that
$\Delta_{+-} = 2\Delta_{\rm SAS}$ at $T_c$.
Thus, the critical temperature is determined by the simultaneous
solution of Eqs.~(\ref{eq:Tc_od}) and (\ref{eq:Ns}) with the
constraint $\Delta_{+-} = 2\Delta_{\rm SAS}$.
Using Eq.~(\ref{eq:n_finiteT}), we find that the $T_c$ equation
may be written after some algebra in the reduced variables
$\beta_c = \Delta_{\rm SAS}^0 / T_c$, $x = N_s / 2 N_0 \Delta_{\rm SAS}^0$,
and $y = N_0 V_{12}$ as
\begin{equation}
2 \beta_c = \ln \left[
  \frac{e^{\beta_c (x y + 1) / 2y} - 1}{e^{\beta_c (x y - 1) / 2y} - 1}
  \right].
\label{eq:Tc}
\end{equation}
The reduced variables $x$ and $y$ are just the axes of the phase diagram
in Fig.~\ref{fig:phase}, in which the contours of constant
$k_B T_c / \Delta_{\rm SAS}^0$ are also shown.
We note that the critical temperature can be of the order of
$\Delta_{\rm SAS}^0 / k_B$, which may be on the order of 10~K for double
quantum wells of the type shown in Fig.~\ref{figure1}.

Below the transition temperature, the order parameter becomes finite,
and we must solve the full non-linear set of equations.
To relate the self-energy parameters obtained in this way to
the staggered spin density in Eq.~(\ref{eq:OP}),
we first note that Eq.~(\ref{eq:Gspec}) can be used to show that
\begin{equation}
\sum_{\bf k} \left< c_{a{\bf k}}^{\dag} c_{b{\bf k}}^{~} \right> =
  \varphi_b^{c} n^{c} \varphi_a^{*c}.
\end{equation}
Inserting this result into Eq.~(\ref{eq:OP}) and
applying Eqs.~(\ref{eq:evec}) and (\ref{eq:uv}), we find that
the staggered magnetization
\begin{equation}
\left[ \rho_{12}^{x} + \rho_{21}^{x} \right] =
  - 2 (n^{-} - n^{+}) \, \frac{\Sigma_{\rm od}}{\Delta_{+-}}.
\label{eq:OP_tn0}
\end{equation}
At zero temperature, this expression reduces to
\begin{equation}
\left[ \rho_{12}^{x} + \rho_{21}^{x} \right] =
  - N_s \, \left[ 1 - 
  \left( \frac{2 \Delta_{\rm SAS}^0}{V_{12} N_s} \right)^2 \right]^{1/2}.
\label{eq:OP_t0}
\end{equation}
Solving the $T > 0$ self-energy and band filling equations for a variety
of interaction strengths and substituting the results into
Eq.~(\ref{eq:OP_tn0}) yields the curves in Fig.~\ref{fig:AFop}.
We see that the staggered magnetization rises rapidly from zero below
$T_c$ and saturates quickly to its $T = 0$ value.
This behavior is generally expected for an order parameter in
mean-field theory.

Another quantity of theoretical and possibly experimental interest
is the specific heat.
The specific heat is proportional to the second derivative of the
free energy with respect to temperature, so this quantity is discontinuous
at either a first- or a second-order phase transition.
This discontinuity is in principle measurable and would provide direct
evidence of a thermodynamic phase transition occurring in these systems.
Actually observing this discontinuity in semiconductor devices of the
kind we are considering would be difficult, however, due to the
low concentration of the relevant electrons.
The specific heat is computed from Eq.~(\ref{eq:Cv}) for a particular
temperature with the eigenenergies and chemical potential obtained
from the self-consistency equations.
Using Eq.~(\ref{eq:Ec}) in Eq.~(\ref{eq:Cv}), the specific heat per
sample area may be written
\begin{equation}
c_V = N_0 T \, \sum_c \int_{-\beta (\mu - e^c)}^{+\infty}
  \frac{x e^x \, dx}{(e^x + 1)^2} \,
  \left( x + \frac{d (\mu - e^c)}{dT} \right),
\label{eq:cv_T}
\end{equation}
which at low temperature reduces to
\begin{equation}
c_V = \frac{\pi^2}{3} N_{\rm occ} N_0 T,
\label{eq:cv_T0}
\end{equation}
where $N_{\rm occ}$ is the number of occupied subbands:
$N_{\rm occ} = 2$ in the $N_1$ and $AF$ phases and
$N_{\rm occ} = 4$ in the $N_2$ phase.
Eq.~(\ref{eq:cv_T0}) is just the usual result obtained from
a Sommerfeld expansion.\cite{Ashcroft}

Since we are aware of no calculations in the literature regarding the specific
heat of a paramagnetic electron gas in a quantum well structure,
we present in Fig.~\ref{fig:cv}(a) $c_V$ for the non-interacting electron
gas to use as a comparison for the interacting case.
At low temperature, the curves naturally resolve themselves into two
groups according to whether one or two subbands are occupied at $T = 0$.
This feature follows directly from Eq.~(\ref{eq:cv_T0}) and is
emphasized in the inset to Fig.~\ref{fig:cv}(a).
Observe that Eq.~(\ref{eq:cv_T0}) is an inadequate description
of $c_V$ when $T$ is larger than only a small fraction of the subband
splitting $\Delta_{\rm SAS}^0$.
The precise fraction is density-dependent, as is whether the actual
specific heat is smaller or larger than Eq.~(\ref{eq:cv_T0}) predicts.
The reason underlying this behavior is that, unlike the metallic case,
the Fermi energy and the temperature are often comparable in these
quantum well structures, invalidating the Sommerfeld expansion.
At higher temperatures, the form of the specific heat for both
one- and two-subband-occupied ground states are similar and have
a magnitude at fixed temperature which increases monotonically with density.
When the temperature becomes of the order of the energies of the
higher subbands, this two-subband description breaks down.
Of course, Eq.~(\ref{eq:cv_T}) may still be used to compute $c_V$
at these temperatures providing the higher subbands are included in
this equation and the equation determining the chemical potential.

With the non-interacting specific heat as a baseline, we can now
examine $c_V$ in the interacting system in both the
normal and antiferromagnetic phases.
Fig.~\ref{fig:cv}(b) presents the specific heat as a function of
temperature calculated for a fixed interaction strength.
The different curves show the evolution of $c_V (T)$ as the density is
increased from the single-subband-occupied $N_1$ phase into the
antiferromagnetic phase.
The specific heat in the $N_1$ phase is similar to the non-interacting
plots in Fig.~\ref{fig:cv}(a), exhibiting the low-temperature, linear-in-$T$
behavior expected from Eq.~(\ref{eq:cv_T0}) which crosses over at some
density-dependent temperature $k_B T \ll \Delta_{\rm SAS}^0$ to an
approximately constant value.
As one enters the antiferromagnetic phase, one sees a discontinuity
develop in $c_V$ at $T_c$, which signals the phase transition.
Unlike superconductivity,\cite{Schrieffer} there is no universal value
for this discontinuity due to the large renormalizing effects of the
temperature on the band structure parameters.
Above $T_c$, the curves resemble their normal-state counterparts,
while below $T_c$, the low-temperature result Eq.~(\ref{eq:cv_T0})
seems to hold.

The latter behavior is not universal, however, as seen from
Fig.~\ref{fig:cv}(c).
This figure shows the evolution of the specific heat function at a
fixed density and increasing interaction.
The system is initially in the two-subband-occupied $N_2$ phase and
becomes antiferromagnetic when $N_0 V_{12} > 1/2$ [cf. Fig.~\ref{fig:phase}].
In the normal state, $c_V (T)$ has the shape expected from
Fig.~\ref{fig:cv}(a) with a low-temperature slope characteristic of
having two subbands occupied (i.e., $N_{\rm occ} = 4$ in Eq.~(\ref{eq:cv_T0})).
Increasing the interaction strength has only minor effects until
the antiferromagnetic region of the phase diagram is reached, at which
point $c_V$ develops a discontinuity at $T_c$.
At the same time, the slope of the low-temperature specific heat drops
by a factor of two in accordance with Eq.~(\ref{eq:cv_T0}) and the fact
that only the lowest interacting bands are occupied in the antiferromagnetic
state.
Above $T_c$, the specific heat is qualitatively similar to the
other normal-state curves.
Below $T_c$, and in contrast to
Fig.~\ref{fig:cv}(b), $c_V \propto T$ at both low temperatures and
temperatures near $T_c$ but with different slopes.
The slope near $T_c$ is a function of the interaction strength and
equals the low-temperature value for $N_0 V_{12} = 1$;
for $N_0 V_{12} > 1$, the slope near $T_c$ is actually less than the
low-temperature slope.

\subsection{Collective Excitations}
\label{sec:AFmodes}

In addition to the ground state and thermodynamic properties of the
antiferromagnetic phase, it is also important to examine its collective
excitations.
The first indication of the antiferromagnetic phase transition
is the disappearance of the intersubband spin-density excitations,
and, on general theoretical grounds, one would like to know what
replaces them in the broken-symmetry phase.
Moreover, experimental studies of semiconductor heterostructures
by inelastic light scattering can measure these excitations, and
theory should provide some guidance about the expected signatures
of the new phase.
The latter point is particularly important in light of current searches
for this phase.\cite{pla-pin}

The basis for studying the collective modes in the antiferromagnetic
phase is Eqs.~(\ref{eq:Pi0_phi}) and (\ref{eq:Pi_cm}) supplemented
by the eigenvalues $\varphi_a^c$, eigenvectors $e^c$, and chemical
potential determined as in the preceding Subsection.
Because the wave functions of the interacting quasiparticles mix
different subbands and spins, the bubble $\Pi_{ab,cd}^{(0)} (q)$
is no longer diagonal in these indices, and so Eq.~(\ref{eq:Pi_cm})
becomes a $16 \times 16$ matrix equation in subband and spin space.
The remaining symmetries in the antiferromagnetic phase do not seem to be
amenable to decomposing this matrix equation and arriving at an analytic
solution, which forces us to adopt a numerical approach.
We therefore obtain the interacting polarizability $\Pi_{ab,cd} (q)$
by direct numerical inversion of Eq.~(\ref{eq:Pi_cm}) at $T = 0$
and identify the collective modes from the condition
\begin{equation}
{\rm det} \, \left[ \delta_{ac} \delta_{bd} - \Pi_{ab,ef}^{(0)} (q) \,
  \left( V_{fe,cd} - V_{fd,ce} \right) \right] = 0.
\label{eq:cm}
\end{equation}
Since we are interested primarily in the intersubband spin-density modes,
we will focus on the intersubband spin-flip polarizability
\begin{equation}
\Pi_{\rm inter} (q) \equiv
  f_{ab}^{\rm ext} \, \Pi_{ab,cd} (q) \, f_{cd}^{\rm ext}
\label{eq:Pi_inter}
\end{equation}
with $f_{ab}^{\rm ext}$ given by Eq.~(\ref{eq:fext}).
Typical results for the spectral function of this polarizability
in the antiferromagnetic phase, $- {\rm Im} \, \Pi_{\rm inter} (q)$,
are shown in Fig.~\ref{fig:qw}.

In analyzing these figures, it is useful to keep in mind the following
facts about the band structure of the antiferromagnetic phase revealed
by the analysis of the preceding Subsection:
(1) the interacting band structure consists of two sets
of two degenerate parabolic subbands separated by an energy $\Delta_{+-}$,
(2) only the lower set of interacting subbands are occupied, yielding
a single Fermi wave vector $k_F$, and
(3) the wave functions corresponding to the interacting bands are a
mixture of the wave functions of the non-interacting subbands.
Facts (1) and (2) indicate that the band structure is similar to that in the
$N_1$ phase, so we expect to see a region of intersubband particle-hole
excitations in the spectral function similar to those in Fig.~\ref{fig:A1-A2}.
The kinematics of these excitations require that their spectral weight
start at $\hbar\omega = \Delta_{+-}$ at $q = 0$ and spread within
the boundaries given by
\begin{equation}
\frac{\hbar^2}{2m^*} \, (q - k_F)^2 \le \hbar\omega - \Delta_{+-} + \mu
  \le \frac{\hbar^2}{2m^*} \, (q + k_F)^2
\end{equation}
for $q > 0$ and $\mu$ measured from the bottom of the lowest interacting
subband.
In the normal phases, these intersubband excitations are cleanly separated
from the intra-subband excitations by the projection in
Eq.~(\ref{eq:Pi_inter}).
In the antiferromagnetic state, however, fact (3) indicates that
such a separation is impossible, leading to the additional low-frequency
particle-hole continuum present in Fig.~\ref{fig:qw}.
Kinematics again show that this region is defined by
$0 < \omega \le \frac{\hbar^2}{2m^*} (q + k_F)^2 - \mu$ for $q < 2 k_F$.
Although not shown in Fig.~\ref{fig:qw}, at larger $q > 2 k_F$, the extent of
this region of particle-hole excitations is defined by
\begin{equation}
\frac{\hbar^2}{2m^*} (q - k_F)^2 - \mu \le
  \omega \le \frac{\hbar^2}{2m^*} (q + k_F)^2 - \mu.
\end{equation}

Within the particle-hole continuum, we find one striking feature
at low frequencies which appears in Fig.~\ref{fig:qw} as a dark,
linearly dispersing feature at low frequencies.
Examining the solutions of Eq.~(\ref{eq:cm}),\cite{cm-solution}
we find that this feature is in fact a Landau-damped collective mode
of the system.
The polarization of this collective mode is extracted by returning
to the full interacting polarizability $\Pi_{ab,cd} (q)$ and
applying Eqs.~(\ref{eq:polarization})-(\ref{eq:fext}) at the
wave vectors and frequencies lying on the dispersion curve
for this mode.
A real-space representation of the resulting spin density displacements
is given in Fig.~\ref{fig:polarization}(b).
To interpret these results, we first note that the antiferromagnetic
phase for these calculations has its spin density oriented along the
$\hat{x}$ direction.
From Fig.~\ref{fig:polarization}(b), we see that the collective mode
corresponds to a wave of antiferromagnetic spin displacements normal
to this orientation, namely in the $\hat{y}$ direction, traveling
transverse to the layering direction of the quantum wells.
The spin displacements are in opposite directions in different wells,
so the net effect is a rotation of the total spin density in the $xy$
plane that preserves the antiferromagnetic correlation of
the spin density between the wells.
Note that there is another collective mode degenerate with this one
which corresponds to a rotation of the total spin density in the $xz$
plane; this mode is projected out by our choice of $f_{ab}^{\rm ext}$
[Eq.~(\ref{eq:fext})].

Outside of the particle-hole continuum, we can look for the undamped
collective modes in the same way as in the preceding
paragraph.\cite{cm-solution}
We find a single optical excitation whose dispersion is indicated by
the thick black line in Fig.~\ref{fig:qw}.
The polarization of this mode is obtained from
Eqs.~(\ref{eq:polarization})-(\ref{eq:fext}) and is shown in
Fig.~\ref{fig:polarization}(c).
From this figure and our knowledge of the orientation of the
spin density in the antiferromagnetic phase, we conclude that
the optical mode corresponds to a modulation of the magnitude of the
spin density which alters neither its direction in space
nor the antiferromagnetic correlation between the wells.
We note that, unlike the low-frequency mode, the optical mode becomes
so strongly Landau damped once it enters the particle-hole continuum
that it is no longer identifiable.

The two collective modes that appear in our calculations can be understood
from general principles of phase transitions involving the
breaking of a continuous symmetry.\cite{NG,Forster}
In our case, the continuous symmetry is spin-rotation or $SU(2)$
invariance, and the extent to which it is broken
is quantified by the staggered spin density ${\bf N}$ [Eq.~\ref{eq:OP_N})].
The collective modes in the broken-symmetry phase correspond
to the modulation in space and time of either the direction or the magnitude
of ${\bf N}$ and are therefore called phase and amplitude modes,
respectively.
The polarization of the collective excitations discussed above
unambiguously identify the two low-frequency, Landau-damped excitations
as phase modes with orthogonal polarizations and the optical excitation
as the amplitude mode.
This identification is strengthened by the dispersion of these excitations:
the phase mode should have an energy which vanishes as $q \rightarrow 0$,
since ground states with different orientations of ${\bf N}$ are degenerate, whereas the amplitude mode should possess an excitation gap.
These expectations are borne out in Fig.~\ref{fig:qw}.

Further insight into these collective modes can be gained
by following the evolution of the intersubband spin-density excitations
as we change the parameters of our model and move from the normal to
antiferromagnetic phases.
We start in the single-subband-occupied $N_1$ phase at the point
marked $A_1$ in Fig.~\ref{fig:phase}.
Increasing the density at fixed interaction, we move towards the
antiferromagnetic phase (point $A_2$), causing the three degenerate
intersubband spin-density excitations to soften as shown in
Fig.~\ref{fig:A1-A2}.
When the $q = 0$ excitations vanish, the system enters the antiferromagnetic
phase and the three intersubband SDEs turn into two degenerate phase
modes and an amplitude mode as seen in Fig.~\ref{fig:qw}(a).
Increasing the density still further (to point $B_2$), the amplitude mode
moves to higher frequencies while the phase mode is largely unchanged
[Fig.~\ref{fig:qw}(b)].

This behavior is typical of entry into the antiferromagnetic phase,
and can be summarized by a plot of the $q = 0$ intersubband excitation
spectrum as a function of the model parameters, Fig.~\ref{fig:w0}.
In Fig.~\ref{fig:w0}(a), we see the transition just described, in
which the system is initially in the $N_1$ phase and the density is
increased at fixed interaction strength.
The collapse of the intersubband spin-density-excitation and the
emergence of the amplitude mode is clearly seen.
We also observe that both the splitting between the interacting subbands
and the intersubband charge-density excitations are continuous and
non-zero across the antiferromagnetic transition, but both have a
discontinuity in their first derivatives.

The antiferromagnetic transition has a slightly different character
when it proceeds from the two-subband-occupied $N_2$ side.
Starting in the $N_2$ phase and increasing the interaction strength
at fixed density, we see from Fig.~\ref{fig:w0}(b) that the
intersubband spin-density excitation softens as before but the
amplitude mode in the antiferromagnetic phase appears immediately
thereafter at finite frequency.
The interacting subband splitting and the intersubband charge-density
excitation are also discontinuous across this phase boundary.
These jumps are a consequence of the first-order nature of the $T = 0$
antiferromagnetic transition from the $N_2$ side.
Although finite temperatures will probably restore continuity to
these curves, the large changes indicated may be experimentally
observable and could provide strong evidence for the transition.

Whether the changes in the intersubband excitation spectrum we have
discussed are observable in, for example, inelastic
light scattering experiments depends to a great extent on the
spectral weight associated with these features.
To give an idea of the range of intensities involved, we plot in
Fig.~\ref{fig:I_B2} a cross section at fixed $q$ of the spectral
weight of the intersubband polarizability from Fig.~\ref{fig:qw}(b).
With increasing frequency, peaks associated with the phase mode, the
amplitude mode, and the intersubband particle-hole continuum are
visible, but the spectral weights associated with each peak are vastly
different with the response dominated by the low-frequency phase mode.
Most inelastic light scattering measurements are done at small $q$ and
moderate frequencies on this scale, so the phase mode may be difficult
to observe unless a concerted effort is made to look for it.
Indeed, the signature of the antiferromagnetic phase in conventional
light scattering experiments may simply be the apparent absence of
all intensity.
We also note that the polarization of the scattered light relative
to the antiferromagnetic ordering direction may affect the observed
intensities of these modes.
Since the ordering direction is arbitrary, this effect may result in
a strong variation in the observed light scattering spectra after
temperature cycling above $T_c$ or between different samples.

\section{DISCUSSION AND SUMMARY}
\label{sec:conclusion}

In this paper, we have studied the magnetic instabilities of
semiconductor quantum wells within the local-density approximation
to density-functional theory and a self-consistent Hartree-Fock theory.
To create a consistent picture of the results of these calculations,
one must realize that these two formalisms supply complementary information.
The LDA computations are designed to be quantitatively reliable for the
normal-state properties of these quantum well structures.
The self-consistent Hartree-Fock calculation, on the other hand,
is only qualitatively reliable, but it is able to describe broken-symmetry
phases that cannot be studied within the LDA.
In particular, there is little point in trying to relate the parameters
from the self-consistent Hartree-Fock calculation to the LDA results,
because the former neglects such real-world effects as the distribution
of the donor impurities which the latter includes.
Hence, the LDA calculations should indicate whether or not the transition
occurs and suggest the structures and densities at which to look for it,
and the self-consistent Hartree-Fock calculations should provide
information on qualitative feature of the resulting antiferromagnetic
phase that can assist experimentalists in identifying it.

The only qualitative point about which the LDA and self-consistent
Hartree-Fock calculations disagree is the ordering wave vector of the
transition from the two-subband-occupied side of the phase diagram:
LDA yields $q_c = k_{F1} - k_{F2}$ while the self-consistent
Hartree-Fock calculation gives $q_c = 0$.
As mentioned in Sec.~\ref{sec:ns_instability}, the discrepancy
between the two formalisms can be traced back to the number of subbands
included in the calculation and not to the form of the interaction,
which is taken to be independent of wave vector in both cases.
Since one expects a calculation including more subbands to be more
accurate, it is reasonable to conclude that at least some part of the
true phase diagram would have $q_c \neq 0$.
That not all of the soft-SDE region of the phase diagram would have
$q_c \neq 0$ is demonstrated by that fact that the one-subband-occupied
SDEs unambiguously soften at $q_c = 0$ in both the LDA and
self-consistent Hartree-Fock theories.
Since the latter theory focuses on inter-well effects and neglects
intra-well ones, its predictions regarding inter-well properties such
as the long-wave-length intersubband spin-density excitations may be
qualitatively valid even in the $q_c \neq 0$ phase.
This phase would have a non-trivial spin-density modulation
transverse to the quantum well layering direction, but whether this
modulation would be of the form of a simple spin-density wave or
something akin to an antiferromagnetic Skyrmion lattice\cite{Brey}
in zero field cannot be determined from the present calculations.
Future investigations exploring the $q_c \neq 0$ phase, and
in particular the nature of the crossover between the $q_c = 0$ and
$q_c \neq 0$ phases, could in principle be performed within a
generalization of the self-consistent Hartree-Fock formalism discussed
in this paper.

With these caveats in mind, let us summarize the primary results of
this work.
We have presented a TDLA calculation which shows
that the intersubband spin-density excitations (SDE) in certain coupled
double- and wide single-quantum-well structures soften completely
in a range of densities around the point where the second subband
begins to populate ($10^{10}-10^{11}$ cm$^{-2}$) and
in the absence of an external magnetic field.
Based on these calculations, we have constructed a phase diagram
indicating the structures likely to exhibit this instability.
We have also computed the excitation spectrum measurable by inelastic
light scattering near the instability in order to illustrate how the
SDE softening would appear in these experiments.
Since the TDLDA yields both spin- and charge-density excitation
spectra which are in very good quantitative agreement with
experiment,\cite{pinczuk,shanabrook,tam-das-94-3} the softening
of the spin-density excitations should be observable
in the appropriate range of densities.

In trying to understand this instability, we have explored the
possibility of ferromagnetic transitions in double- and single-quantum-well
structures by including the spin degree of freedom in a
density-functional calculation within the LSDA.
We find that a ferromagnetic transition occurs in the double-quantum-well
structures which exhibit the SDE softening but that the transition occurs
at much lower ($\sim 10^{9}$ cm$^{-2}$) densities, implying
that the SDE softening cannot be associated with ferromagnetism.
In square single quantum wells, our computations provide evidence
for a spin-polarized phase of the electron gas which lies between
the Wigner crystal and normal phases.
The critical density for this transition decreases with increasing well
width, demonstrating that exchange-correlation effects are stronger in
lower dimensions, as expected from a simple Hartree-Fock analysis.

Having failed to identify the SDE-softened phase within density-functional
theory, we turned to a simple model of coupled double quantum wells
which we treated within self-consistent Hartree-Fock theory.
This model is able to reproduce the SDE softening in its normal state,
and the polarization of the soft mode indicates that the softening
signals the onset of antiferromagnetic order in the spin density
between the wells.
Extending our calculations into the antiferromagnetic phase, we
find that this phase exists and is stable over a wide range of
parameters and that the mean-field transition temperature
can be of the same order as the symmetric-antisymmetric splitting.
In addition, we find that the transition to this phase at
zero temperature is second-order from the single-subband-occupied side
of the phase diagram, but first-order from the two-subband-occupied side.
Due to the absence of an energy gap in the single-particle spectrum, we
do not expect strong anomalies in the transport properties to accompany
the transition; however, our calculations of the electronic specific heat
show that, if this quantity is measurable, it will show a characteristic
discontinuity at the transition temperature.
A means of searching for the transition which is more likely to
succeed is the measurement of the collective spin-density excitations
through inelastic light scattering.
By computing the spectrum of these excitations in the antiferromagnetic phase,
we identify a Landau-damped phase mode of the order parameter and
a true optical collective excitation corresponding to the amplitude mode.
The spectral weight associated with the phase mode is large, suggesting
that inelastic light scattering experiments should look at low frequencies
for this characteristic excitation of the antiferromagnetic phase.

In closing, we note that at least one experimental group has investigated
the possibility of an antiferromagnetic phase transition of the
type we describe by performing resonant inelastic light scattering
measurements on double-quantum-well structures.\cite{pla-pin}
The results, however, have been mixed.
In zero field, the complete softening of the spin-density excitation
does not seem to appear in the electron density regime
($\approx 5 \times 10^{10}$ cm$^{-2}$) predicted by the TDLDA theory.
It is possible, of course, that impurity-scattering-induced broadening
effects make it impossible to observe the complete softening of the
spin-density excitation.
Alternatively, the TDLDA theory may overestimate the density range
in which the transition occurs, implying that the actual antiferromagnetic
instability may take place at lower electron densities.
A third possibility is that the TDLDA approach may simply be
inadequate for studying semiconductor quantum wells at the low
densities involved.

More promising are the experimental results in small but finite
magnetic fields along the layering direction, which do indicate a
softening of the intersubband spin-density excitations at the filling
factors $\nu = 2$ and 6.\cite{pla-pin}
The general observation of the softening of the intersubband
spin-density excitations in the presence of a magnetic field\cite{pla-pin}
is qualitatively consistent with the prediction of our zero-field theory,
since a magnetic field weak enough so that the system in not completely
spin-polarized enhances the effects of the interaction by
reducing the kinetic energy through Landau quantization.
Thus, the magnetic field naturally enhances the potential for the
type of spin instabilities discussed in this paper to appear.
Moreover, the basic Hartree-Fock theory underlying the description of
the resulting broken-symmetry states do not change, although the
effects of Landau-level quantization should be included.
Such a generalization of the self-consistent Hartree-Fock approach
is straightforward and is left for future work.

\section*{ACKNOWLEDGMENTS}

The authors would like to thank R.~Decca and P.~B.~Littlewood
for useful discussions.
This work is supported by the US-ARO and the US-ONR.

\twocolumn

\begin{figure}
\caption{Typical coupled double-quantum-well structure, and its 
self-consistent LDA subband energy levels $E_i$, eigenfunctions $\phi_i$, 
electron density $n(z)$, and Fermi energy $E_F$. 
Also shown are the effective, Hartree, and exchange-correlation potentials 
$V_{EFF}$, $V_{H}$, and $V_{XC}$.
The sheet density $N_s=2.68 \times 10^{11}$ cm$^{-2}$.
\label{figure1}}
\end{figure}

\begin{figure}
\caption{Many-body diagrams used to compute (a) the self-energy $\Sigma$,
(b) the contribution of the interactions to the energy
$E_{\rm int} = \left< H_{\rm int} \right>$, (c) the generalized
polarizability $\Pi$, and (d) the vertex function $\gamma$ within
the self-consistent Hartree-Fock approximation.
The solid lines represent dressed electronic propagators
[Eq.~(\protect\ref{eq:G})] and the dashed lines the effective
interaction $V$ [Eq.~(\protect\ref{eq:V})], both of which are
matrices in subband and spin space.
In order to treat both spin and charge polarizabilities with the
same equations, the polarizability is not separated into
reducible and irreducible parts.}
\label{fig:diagrams}
\end{figure}

\begin{figure}
\caption{Calculated intersubband charge-density excitation $E_{\text{CDE}}$, 
spin-density excitation $E_{\text{SDE}}$, and single-particle excitation 
$E_{\text{SPE}} \equiv \bigtriangleup_{\text{SAS}}$
energies as functions of the 2D electron density $N_{\text{S}}$
for a DQW structure with barrier width $d_B$ = 40~\AA and well width 
$d_W$ = 139~\AA.
The critical density for the instability 
$N_{\text{C}} \approx 0.69 \times10^{11}$ cm$^{-2}$.
The bottom figure shows an expanded density range making obvious
the re-entrance of the normal phase at very low electron density.
\label{figure2}}
\end{figure}

\begin{figure}
\caption{Dependence on the sheet density $N_S$ of
(a) the mean-field vertex correction $\mid U_{\rm XC 12} \chi_{12}^0 \mid$
(solid line) and the absolute value of the lowest-order polarizability
$\mid \chi^0_{12} \mid$ (dashed line) and (b) the spin-polarized
exchange-correlation-induced vertex correction $U_{\rm XC 12}$
for the double-quantum-well structure in Fig.~\protect\ref{figure1}.
Note that the electron gas is unstable the range of densities where
$\mid U_{xc} \chi_{12}^0 \mid \geq 1$ (see text).
\label{figure3}}
\end{figure}

\begin{figure}
\caption{Calculated zero-temperature phase diagram for double
quantum wells in terms of the sheet density $N_S$ and the barrier
width $d_B$ for fixed well widths $d_W$ = 139~\AA.
Inset: phase diagram for fixed $d_B$ = 40~\AA in terms of the sheet
density and well widths.
The normal (N) and the triplet excitonic (E) phases are shown.
\label{figure4}}
\end{figure}

\begin{figure}
\caption{Calculated zero-temperature phase diagram for coupled
double quantum wells in terms of $r_s^{2D} \equiv (\pi N_S)^{-1/2} / a_0$
and the dimensionless symmetric-antisymmetric subband
splitting $\Delta_{SAS} / (e^2 / \epsilon (d_W + d_B))$,
where $a_0$ and $\epsilon$ are the Bohr radius and dielectric constant
for GaAs and $d_B$ ($d_W$) is the barrier (well) width.
Solid circles correspond to $d_W$ = 139~\AA and various $d_B$ and are
taken from the main part of Fig.~\protect\ref{figure4}, while the
crosses correspond to $d_B$ = 40~\AA and various $d_W$ and are obtained
from the inset to Fig.~\protect\ref{figure4}.
The normal (N) and the triplet excitonic (E) phases are shown.
\label{figure5}}
\end{figure}

\begin{figure}
\caption{Energy of the intersubband spin-density excitations $E_{\rm SDE}$
as a function of wave vector $q$ in a coupled double-quantum-well system
with a barrier width of 40~\AA and well widths of 139~\AA for
sheet densities $N_S$ in units of $10^{11}$ cm$^{-2}$ approaching the
critical density $N_C \simeq 0.686 \times 10^{11}$ cm$^{-2}$  from above
(thick lines).
The thin lines show the lower boundary of the particle-hole
continuum, above which the collective excitations are Landau damped.
\label{figure6}}
\end{figure}

\begin{figure}
\caption{Calculated Raman scattering spectra in the cross-polarization
geometry for a double-quantum-well structure with a 40~\AA barrier width
and 139~\AA well widths.
The curves illustrate the signatures of the excitonic instability
(a) as the sheet density $N_S$ is lowered to the critical density and
(b) as the wave-vector transfer $q$ is increased at constant sheet
density $N_S=0.7 \times 10^{11}$ cm$^{-2}$.
\label{figure7}}
\end{figure}

\begin{figure}
\caption{Typical wide square quantum well given by the bare
confining potential $V_{CONF}$, and its self-consistent LDA
subband energy levels $E_n$, eigenfunctions $\phi_n$,
electron density $n(z)$, Fermi energy $E_F$,
and effective, Hartree, and exchange-correlation potentials $V_{EFF}$,
$V_{H}$, and $V_{XC}$.
The sheet density is $N_s=0.9 \times 10^{11}$ cm$^{-2}$.
The figure shows how the electronic density profile becomes
localized on the sides of the well, similar to the profile
in a double quantum-well system. Bottom: lowest energies in 
expanded scale.
\label{figure8}}
\end{figure}

\begin{figure}
\caption{Calculated intersubband charge-density excitation energy $E_{CDE}$, 
spin-density excitation energy $E_{SDE}$, and single-particle excitation
energy $E_{SPE} \equiv \Delta_{SAS}$ as functions of the 2D electron 
density $N_{S}$ for the wide square-well structure shown in
Fig.~\protect\ref{figure8}.
Note the collapse of $E_{\rm SDE}$ for
$N_S \simeq 0.2-0.4 \times 10^{11}$ cm$^{-2}$.
\label{figure9}}
\end{figure}

\begin{figure}
\caption{Approximate zero-temperature, spin-polarization phase
diagram of single square wells calculated in the local-spin-density
approximation in terms of the sheet density $N_S$ and well width $d_B$
(top) and the $r_s$ parameter and well width (bottom).
\label{figure10}}
\end{figure}

\begin{figure}
\caption{Mean-field phase diagram of the antiferromagnetic sector
of the point-contact model described in the text.
The independent variables are the normalized intersubband interaction
matrix element $N_0 V_{12}$ and the sheet density
$N_s / 2 N_0 \Delta_{\rm SAS}^0$, where $N_0 = m^* / 2 \pi \hbar^2$
in the single-spin density of states and $\Delta_{\rm SAS}^0$ is
the splitting between the lowest two subbands when $V_{12} = 0$.
The other interaction matrix elements $V_{11}$ = $V_{22}$ = 0.
The regions correspond to the normal (paramagnetic) phase with one
subband occupied ($N_1$), the normal phase with both subbands occupied
($N_2$), and the antiferromagnetic phase (AF).
Contours in the antiferromagnetic region of the phase diagram are
the computed values of the critical temperature $T_c$ for the
antiferromagnetic transition in units of $\Delta_{SAS}^0 / k_B$.
Observe that $k_B T_c$ can be larger than $\Delta_{SAS}^0$, indicating
that the antiferromagnetic phase may persist to observable temperatures.
The other labels in the figure identify points for future reference.}
\label{fig:phase}
\end{figure}

\begin{figure}
\caption{Dispersion of the intersubband spin-density excitations (SDEs)
as the sheet density $N_s$ approaches the antiferromagnetic phase
from the low-density, one-subband-occupied side computed in the
antiferromagnetic sector of the point-contact interaction model
discussed in the text with $N_0 V_{12} = 1.0$.
The thick lines show the energy $\hbar \omega$ of the SDEs
in units of the renormalized
splitting of the lowest two subbands $\Delta_{\rm SAS}^*$ as a
function of wave vector $q$ relative to the Fermi wave vector $k_F$.
The thin lines show the boundaries of the particle-hole continuum,
within which the collective excitations are damped.
The sheet densities and corresponding points in the phase diagram of
Fig.~\protect\ref{fig:phase} are given in the figure.
The transition to the antiferromagnetic phase occurs when the
$q = 0$ SDEs soften at $N_s / 2 N_0 \Delta_{\rm SAS}^0$ = 1.0.
A similar softening appears in Fig.~\protect\ref{figure6}, which
shows the approach to the antiferromagnetic phase from the
high-density side computed within the LDA.}
\label{fig:A1-A2}
\end{figure}

\begin{figure}
\caption{Polarization of (a) an intersubband spin-density excitation
in the normal phase and of the (b) phase (Nambu-Goldstone) and
(c) amplitude modes in the antiferromagnetic phase with the spin
density oriented along the $\hat{x}$ direction computed as described
in the text.
The configuration of the quantum wells is as in
Fig.~\protect\ref{figure1}.
The two planes are sections through this geometry normal to the
layering direction and are located in the center of each well.
The distances in these planes are measured in units of the wavelength
of the collective excitation $\lambda$, whose propagation is in
the $\hat{x}$ direction.
The arrows show the direction and magnitude of the spin density
modulation induced by the collective excitations.
These modes have the form of a traveling wave, so the spin modulation
at a different time is obtained by shifting these pictures along
the $\hat{x}$ direction.
Since the total spin density is the sum of the antiferromagnetic
polarization and the modulations shown in (b) and (c), the identification
of these modes with the phase and amplitude motions is apparent.}
\label{fig:polarization}
\end{figure}

\begin{figure}
\caption{Staggered spin density $(\rho_{12}^x + \rho_{21}^x)$, which
is the order parameter for the antiferromagnetic phase transition
discussed in the text [cf. Eq.~(\protect\ref{eq:OP_N})], normalized
by the electron sheet density $N_s$ as a function of temperature $T$
in units of the splitting between the two lowest subbands in the
non-interacting limit $\Delta_{\rm SAS}^0$.
The curves are computed for $N_s / 2 N_0 \Delta_{\rm SAS}^0 = 2.5$ and
$N_0 V_{12}$ = 0.55 to 0.80 in increments of 0.05;
the lower and upper values correspond to points $D_1$ and $D_2$ in
the phase diagram of Fig.~\protect\ref{fig:phase}.
Inset:  Expectation value of the spin density $\left< s^x (z) \right>$
in real space as a function of the distance along the layering direction
$z$ for the double quantum well of Fig.~\protect\ref{figure1} in the
antiferromagnetic phase.
Note that $\left< s^x (z) \right>$ = 0 in the paramagnetic phase.}
\label{fig:AFop}
\end{figure}

\begin{figure}
\caption{Electronic specific heat at constant volume $c_V$ in normalized by
$2 \pi^2 N_0 \Delta_{\rm SAS}^0 / 3$ as a function
of the temperature $T$ in units of $\Delta_{\rm SAS}^0 / k_B$ for
(a) $N_0 V_{12} = 0$ and $N_s / 2 N_0 \Delta_{\rm SAS}^0$ = 0.1 to
2.0 in increments of 0.1,
(b) $N_0 V_{12} = 1$ and $N_s / 2 N_0 \Delta_{\rm SAS}^0$ = 0.5 to
1.5 in increments of 0.1 (i.e., along the line from $A_1$ to $B_2$
in Fig.~\protect\ref{fig:phase}), and
(c) $N_s / 2 N_0 \Delta_{\rm SAS}^0$ = 2.5 and $N_0 V_{12}$ = 0.2
to 0.8 in increments of 0.05 (i.e., along the line from $C_1$ to
$D_2$ in Fig.~\protect\ref{fig:phase}).
The inset in (a) enlarges the low-temperature portion of the main figure
in order to see the deviation from the analytic low-$T$ expression,
Eq.~(\protect\ref{eq:cv_T0}).
Entrance into the antiferromagnetic phase is signaled by a
discontinuity in the specific heat that is apparent in (b) and (c) but
absent in (a).}
\label{fig:cv}
\end{figure}

\begin{figure}
\caption{Dispersion of the intersubband spin density collective modes
in the antiferromagnetic phase at the points (a) $B_1$ and (b) $B_2$
of the phase diagram of Fig.~\protect\ref{fig:phase}, corresponding
to ($N_s / 2 N_0 \Delta_{\rm SAS}^0$, $N_0 V_{12}$) = (1.1,1.0)
and (1.5,1.0), respectively.
The thick lines are the energy $\hbar \omega$ in units of the non-interacting
subband splitting $\Delta_{SAS}^0$ of the amplitude mode of the
antiferromagnetic order parameter as a function of the wave vector $q$
in units of $q_{\Delta}^2 = m^* \Delta_{\rm SAS}^0 / \hbar^2$.
The shaded region is the particle-hole continuum, with darker shades
representing larger spectral weight than lighter shades on a logarithmic
intensity scale.
The dark linear feature is the phase or Nambu-Goldstone mode of the
order parameter.
Note that this mode is damped by particle-hole excitations and that
intra-subband excitations enter into the spectrum due to the mixing
of the non-interacting wave functions in the symmetry-broken phase.}
\label{fig:qw}
\end{figure}

\begin{figure}
\caption{Normalized energy $\hbar \omega / \Delta_{\rm SAS}^0$
of the $q = 0$, $T = 0$ collective excitations and interacting
subband splitting for (a) a fixed interaction $N_0 V_{12} = 1.0$ and
varying sheet density $N_s / 2 N_0 \Delta_{\rm SAS}^0$ and
(b) varying interaction $N_0 V_{12}$ and fixed sheet density
$N_s / 2 N_0 \Delta_{\rm SAS}^0$ = 2.5 in the antiferromagnetic
sector of the point-contact model discussed in the text.
Illustrated are the intersubband spin-density (SDE, solid lines)
and charge density excitations (CDE, dotted lines), the renormalized
subband splitting $\Delta_{\rm SAS}^*$ (SPE, dot-dashed lines), and the
amplitude mode in the antiferromagnetic phase (dashed line).
The top axes show the corresponding points in the phase diagram in
Fig.~\protect\ref{fig:phase}.
As seen in (b), the collective mode energies are discontinuous across
the two-subband-occupied ($N_2$) to antiferromagnetic phase boundary
at $T = 0$, indicating a first-order transition.}
\label{fig:w0}
\end{figure}

\begin{figure}
\caption{Imaginary part of the intersubband spin-flip polarizability
$-{\rm Im} \, \Pi_{\rm inter}$ [Eq.~(\protect\ref{eq:Pi_inter})]
as a function of the excitation energy
$\hbar \omega$ relative to the non-interacting subband splitting
$\Delta_{\rm SAS}^0$ for $q = 0.4 q_{\Delta}$ at
the point $B_2$ in the antiferromagnetic region of the phase diagram
of Fig.~\protect\ref{fig:phase} [see also Fig.~\protect\ref{fig:qw}(b)].
This quantity is related to the intensity of the intersubband
response in inelastic light scattering experiments.
The phase or Nambu-Goldstone mode dominates the low-frequency
response, the amplitude mode shows much less intensity, and the
particle-hole excitations have a very weak signal.
This spectrum is computed using a finite scattering rate
$\gamma = 0.01 \Delta_{\rm SAS}^0$ to simulate the effects of impurities.}
\label{fig:I_B2}
\end{figure}


\begin{references}

\bibitem[*]{paddress}Present address:  Institut f\"ur Theoretische Physik, 
J.\ W.\ Goethe Universit\"at Frankfurt,
Robert-Mayer-Str.\ 8, D-60054 Frankfurt a.M., Germany.

\bibitem{tsu-sto-gos} D.\ C.\ Tsui, H.\ L.\ St\"{o}rmer, and A.\ C.\ Gossard,
Phys.\ Rev.\ Lett.\ {\bf 48}, 1559 (1982).

\bibitem{lau} R.\ B.\ Laughlin, Phys.\ Rev.\ Lett.\ {\bf 50}, 1395 (1983).

\bibitem{fer} H.\ A.\ Fertig, Phys.\ Rev.\ B {\bf 40}, 1087 (1989);
A.\ H.\ MacDonald, P.\ M.\ Platzman, and G.\ S.\ Boebinger, 
Phys.\ Rev.\ Lett.\ {\bf 65}, 775 (1990);
L.\ Brey, {\it ibid.}, 903 (1990);
X.\ Chen and J.\ J.\ Quinn, {\it ibid.} {\bf 67}, 895 (1991);
S.~He, X.~C.~Xie, S.~Das Sarma, and F.~C.~Zhang,
Phys.\ Rev.\ B {\bf 43}, 9339 (1991).

\bibitem{boe} G.~S.~Boebinger, H.~W.~Jiang, L.~N.~Pfeiffer, and K.~W.~West,
Phys. Rev. Lett. {\bf 64}, 1793 (1990).

\bibitem{eis}
Y.~W.~Suen, L.~W.~Engel, M.~B.~Santos, M.~Shayegan, and D.~C.~Tsui,
Phys.\ Rev.\ Lett.\ {\bf 68}, 1379 (1992);
J.~P.~Eisenstein, G.~S.~Boebinger, L.~N.~Pfeiffer, K.~W.~West, and S.~He,
{\it ibid.}, 1383 (1992).

\bibitem{pri-pla-he} R.\ J.\ Price, P.\ M.\ Platzman, and Song He,
Phys.\ Rev.\ Lett.\ {\bf 70}, 339 (1993).

\bibitem{he-das-xie} S.\ He, S.\ Das Sarma, and X.\ C.\ Xie,
Phys.\ Rev.\ B {\bf 47}, 4411 (1993), and references therein.

\bibitem{hal-lee-rea} B.\ I.\ Halperin, P.\ A.\ Lee, and N.\ Read,
Phys.\ Rev.\ B {\bf 47}, 7312 (1993).

\bibitem{wig} E.\ P.\ Wigner, Phys.\ Rev.\ {\bf 46}, 1002 (1934);
Trans. Faraday Soc. {\bf 34}, 678 (1938).

\bibitem{Grimes}C. C. Grimes and G. Adams,
Phys. Rev. Lett. {\bf 42}, 795 (1979).

\bibitem{spi-den-wav}
S.\ F.\ Edwards and A.\ J.\ Hillel, J.\ Phys.\ C {\bf 1}, 61 (1968);
A.\ W.\ Overhauser, Phys.\ Rev.\ Lett.\ {\bf 3}, 414 (1959);
{\it ibid.} {\bf 4}, 462 (1960);
Phys.\ Rev.\ {\bf 128}, 1437 (1962).

\bibitem{blo} F.\ Bloch, Z.\ Phys.\ {\bf 57}, 549 (1929).

\bibitem{MacDonald}A. H. MacDonald, Phys. Rev. B {\bf 37}, 4792 (1988).

\bibitem{Varma}C. M. Varma, A. I. Larkin, and E. Abrahams,
Phys. Rev. B {\bf 49}, 13999 (1994).

\bibitem{pinczuk} 
A.\ Pinczuk, S.\ Schmitt-Rink, G.\ Danan, J.\ P.\ Valladares, 
L.\ N.\ Pfeiffer, and K.\ W.\ West, Phys.\ Rev.\ Lett. {\bf 63}, 
1633 (1989);
S.\ L.\ Chuang, M.\ S.\ C.\ Luo, S.\ Schmitt-Rink, and
A.\ Pinczuk, Phys.\ Rev.\ B {\bf 46}, 1897 (1992).

\bibitem{shanabrook} 
D.\ Gammon, B.\ V.\ Shanabrook, J.\ C.\ Ryan, and D.\ S.\ Katzer,
Phys.\ Rev.\ B {\bf 41}, 12311 (1990);
D.\ Gammon, B.\ V.\ Shanabrook, J.\ C.\ Ryan, D.\ S.\ Katzer, and
M.\ J.\ Yang, Phys.\ Rev.\ Lett.\ {\bf 68}, 1884 (1992).

\bibitem{Ruden}P. P. Ruden and Z. Wu,
Appl. Phys. Lett. {\bf 59}, 2165 (1991).

\bibitem{nei}L.~\'{S}wierkowski, D.~Neilson, and J.~Szyma\'{n}ski,
Phys.\ Rev.\ Lett.\ {\bf 67}, 240 (1991);
D.~Neilson, L.~\'{S}wierkowski, J.~Szyma\'{n}ski, and L.~Liu,
{\it ibid.} {\bf 71}, 4035 (1993).

\bibitem{mar-das} 
S.\ Das Sarma and I.\ K.\ Marmorkos, Phys.\ Rev.\ B {\bf 47}, 16343 (1993);
I.\ K.\ Marmorkos and S.\ Das Sarma, {\it ibid.} {\bf 48}, 1544 (1993);
and references therein.

\bibitem{Katayama}Y. Katayama, D. C. Tsui, H. C. Manoharan, and M. Shayegan,
Surf. Sci. {\bf 305}, 405 (1994); Phys. Rev. B {\bf 52}, 14817 (1995);
and X. Ying, S. R. Parihar, H. C. Manoharan, and M. Shayegan,
{\it ibid.} {\bf 52}, 11611 (1995).

\bibitem{decca} R.\ Decca, A.\ Pinczuk, S.\ Das Sarma, B.\ S.\ Dennis,
L.\ N.\ Pfeiffer, and K.\ W.\ West, Phys.\ Rev.\ Lett.\, {\bf 72},
1506 (1994).

\bibitem{tam-das-94-3} P.\ I.\ Tamborenea and S.\ Das Sarma,
Phys.\ Rev.\ B {\bf 49}, 16821 (1994).

\bibitem{ern-gon-sya} S.\ Ernst, A.\ R.\ Go\~{n}i, K.\ Syassen, and
K.\ Eberl, Phys.\ Rev.\ Lett.\ {\bf 72}, 4029 (1994).

\bibitem{das-tam-94-4}
S.\ Das Sarma and P.\ I.\ Tamborenea, Phys.\ Rev.\ Lett.\
{\bf  73}, 1971 (1994).

\bibitem{rad-das} R.~J.~Radtke and S.~Das Sarma,
Solid State Commun. {\bf 96}, 215 (1995); {\it ibid.} (to appear).

\bibitem{exciton-note}As we shall see later, it is actually somewhat
misleading to think about the spin-density-excitation condensate as
a sea of bound electron-hole pairs from different subbands.

\bibitem{ando76} T.\ Ando, Surf.\ Sci.\ {\bf 58}, 128 (1976);
T.\ Ando, Phys.\ Rev.\ B {\bf 13}, 3468 (1976).

\bibitem{kat-and} T.\ Ando, J.\ Phys.\ Soc.\ Jpn. {\bf 51}, 3893 (1982);
S.\ Katayama and T.\ Ando, {\it ibid.} {\bf 54}, 1615 (1985).

\bibitem{pin-abs} A.\ Pinczuk and G.\ Abstreiter, 
in {\it Light Scattering in Solids V}, edited by M.\ Cardona and G.\
Guntherodt (Springer-Verlag, Berlin, 1989).

\bibitem{das:ils} S.\ Das Sarma, in 
{\it Light Scattering in Semiconductor Structures and Superlattices}, 
edited by D.\ J.\ Lockwood and J.\ F.\ Young (Plenum Press, New York, 1991).

\bibitem{hoh-koh} P.\ Hohenberg and W.\ Kohn, Phys.\ Rev.\,
{\bf 136}, B864 (1964).

\bibitem{koh-sha} W.\ Kohn and L.\ J.\ Sham, Phys.\ Rev.\,
{\bf 140}, A1133 (1965).

\bibitem{ste-das} F.\ Stern and S.\ Das Sarma, Phys.\ Rev.\ B {\bf 30},
840 (1984).

\bibitem{cep-ald}
D.\ M.\ Ceperley and B.\ J.\ Alder, Phys.\ Rev.\ Lett.\ {\bf 45}, 566 (1980);
J.\ Perdew and A.\ Zunger, Phys.\ Rev.\ B {\bf 23}, 5048 (1981).

\bibitem{ham-mcw} D.\ C.\ Hamilton and A.\ L.\ McWhorther,
in {\it Proceedings of the International Conference on Light Scattering
Spectra in Solids} (Springer-Verlag, Berlin, 1969), p.~309.

\bibitem{fet-wal}A.\ L.\ Fetter and J.\ D.\ Walecka,
{\it Quantum Theory of Many-Particle Systems}, (McGraw-Hill, New York, 1971).

\bibitem{sds-light-scatt}
S.\ Das Sarma, in {\it Light Scattering in Semiconductor
Structures and Superlattices}, edited by D.\ J.\ Lockwood and J.\ F.\ Young
(Plenum Press, New York, 1991), and references therein.

\bibitem{jai-das} J.\ K.\ Jain and S.\ Das Sarma, Phys.\ Rev.\ B
{\bf 36}, 5949 (1987).

\bibitem{von-hed} U.\ von Barth and L.\ Hedin, J.\ Phys.\ C\ {\bf 5},
1629 (1972).

\bibitem{pan-raj} M.\ M.\ Pant and A.\ K.\ Rajagopal, Solid State Commun.\
{\bf 10}, 1157 (1972).

\bibitem{dre-gro}R.\ M.\ Dreizler and E.\ K.\ U.\ Gross,
{\it Density Functional Theory} (Springer-Verlag, Berlin, 1990).

\bibitem{Rajagopal}A~.K.~Rajagopal and J.~C.~Kimball,
Phys. Rev. B {\bf 15}, 2819 (1977).

\bibitem{cep-unp} D.\ M.\ Ceperley, private communication.


\bibitem{tan-cep} B.\ Tanatar and D.\ M.\ Ceperley, 
Phys.\ Rev.\ B {\bf 39}, 5005 (1989).

\bibitem{das-vin:val-deg}
S.\ Das Sarma and B.\ Vinter, Phys.\ Rev.\ B {\bf 26}, 960 (1982);
{\em ibid.} {\bf 28}, 3639 (1983).

\bibitem{hem-mas-kwi} C.\ E.\ Hembree, B.\ A.\ Mason, J.\ T.\ Kwiatkowski,
J.\ Furneaux, and J.\ A.\ Slinkman, Phys.\ Rev.\ B {\bf 48}, 9162 (1993).

\bibitem{Schrieffer}J.~R.~Schrieffer, {\it Theory of Superconductivity}
(Addison-Wesley, New York, 1964).

\bibitem{Mahan}G.~D.~Mahan, {\it Many-Particle Physics}
(Plenum, New York, 1981).

\bibitem{Forster}See, for example, D.~Forster,
{\it Hydrodynamic Fluctuations, Broken Symmetry, and Correlation Functions}
(Addison-Wesley, New York, 1975).

\bibitem{Baym}G.~Baym and L.~P.~Kadanoff,
Phys. Rev. {\bf 124}, 287 (1961);
G.~Baym, {\it ibid.} {\bf 127}, 1391 (1962).

\bibitem{footnote} We can obtain a crude estimate of the upper limit to the 
transition temperature $T_c$ by putting
$k_B T_c \approx \bigtriangleup_{\text{SAS}}$, which makes the TDLDA
calculations meaningful only for
$T \ll T_c \approx \bigtriangleup_{\text{SAS}}/k_B$.
For $T \approx$ 1-4~K, this restricts $d_B \le$ 45-50~\AA.

\bibitem{gun-lun}
O.\ Gunnarsson and B.\ I.\ Lundqvist, Phys.\ Rev.\ B {\bf 13}, 4274 (1976).

\bibitem{Rice}For a review, see B.~I.~Halperin and T.~M.~Rice,
in {\it Solid State Physics}, Vol.~21, edited by F.~Seitz, D.~Turnbull,
and H.~Ehrenreich (Academic, New York, 1968), p.~115.

\bibitem{num-note}This {\it ansatz} is confirmed by numerical solution
of the complete self-energy equations with the band-filling constraint.

\bibitem{Ashcroft}N.~W.~Ashcroft and N.~D.~Mermin, {\it Solid State Physics}
(Saunders College, Philadelphia, 1976), Chap.~1.

\bibitem{pla-pin}A.~S.~Plaut, A.~Pinczuk, B.~S.~Dennis, J.~P.~Eisenstein,
L.~N.~Pfeiffer, and K.~W.~West, Surf. Sci. (to appear);
A.~S.~Plaut, Bull. Am. Phys. Soc. {\bf 41}(1), 590 (1996);
and A.~S.~Plaut (private communication).

\bibitem{cm-solution}For technical reasons, it actually is more
convenient to find the collective modes in the intersubband spin-density
channel by locating the points where $1 / {\rm Re}\,\Pi_{\rm inter} (q) = 0$
rather than solving Eq.~(\ref{eq:cm}) directly.
The two methods yield identical results for the intersubband SDE modes.

\bibitem{NG}Y. Nambu and G. Jona-Lasinio, Phys. Rev. {\bf 122}, 345 (1961);
J. Goldstone, Nuovo Cimento {\bf 19}, 154 (1961).

\bibitem{Brey}L.~Brey, H.~A.~Fertig, R.~Cot\'{e}, and A.~H.~MacDonald,
Phys. Rev. Lett. {\bf 75}, 2562 (1995).

\end{references}
\end{document}